\documentclass{aa}  

\usepackage{graphicx}
\usepackage{txfonts}
\usepackage[colorlinks=true,citecolor=blue,linkcolor=blue]{hyperref}
\usepackage{subfigure}
\usepackage{subcaption}

\bibpunct{(}{)}{;}{a}{}{,}
\defcitealias{andersson24starclusters}{A24}

\usepackage{orcidlink}

\usepackage{etoolbox}
\makeatletter
\AtBeginDocument{
  \@ifpackageloaded{lineno}{\nolinenumbers\renewcommand{\linenumbers}{} }{}
}
\makeatother

\begin{document}

   \title{Resolving the thermal component of stellar feedback with the "anchor-and-refine" scheme in galactic simulations}

\titlerunning{Resolving thermal feedback}
\authorrunning{Chengzhe~Li et al.}
   \author{Chengzhe Li\inst{1}\orcidlink{0000-0002-0480-2276}
        \and Hui Li\inst{1}\fnmsep\thanks{E-mail: hliastro@tsinghua.edu.cn}\orcidlink{0000-0002-1253-2763}
        \and Wei Cui\inst{1}\orcidlink{0000-0002-6324-5772}
        \and Federico Marinacci\inst{2,3}\orcidlink{0000-0003-3816-7028}
        \and Laura V. Sales\inst{4}\orcidlink{0000-0002-3790-720X}
        \and \\ Mark Vogelsberger\inst{5,6}\orcidlink{0000-0001-8593-7692} 
        \and Paul Torrey\inst{7}\orcidlink{0000-0002-5653-0786}
          }
   \institute{Department of Astronomy, Tsinghua University, Beijing 100084, China
   \and Department of Physics \& Astronomy ``Augusto Righi", University of Bologna, via Gobetti 93/2, 40129 Bologna, Italy
   \and INAF, Astrophysics and Space Science Observatory Bologna, Via P. Gobetti 93/3, I-40129 Bologna, Italy
   \and Department of Physics and Astronomy, University of California, Riverside, CA, 92521, USA
   \and Department of Physics \& Kavli Institute for Astrophysics and Space Research, Massachusetts Institute of Technology, Cambridge, MA 02139, USA
   \and Fachbereich Physik, Philipps Universit\"at Marburg, D-35032 Marburg, Germany
   \and Department of Astronomy, University of Virginia, Charlottesville, VA 22904, USA}

   \date{Received ; accepted }
 
  \abstract  
   {Stellar feedback plays a crucial role in regulating baryon cycles of a galactic ecosystem. However, galaxy-scale simulations still do not have high enough resolutions (typically a gas cell/particle has mass $m_{\rm i, gas}> 10^{3} \, \rm{M_\odot}$) to resolve the cooling radius. Sometimes a momentum-flavored subgrid model for feedback is used in coarse resolutions. This treatment of feedback poses an issue that the important hot gas around stars is not or under resolved.}
   {We aim to develop a robust, resolution-adaptive, stellar feedback model that focus on capturing the thermal properties of feedback in galactic simulations, which performs robustly at a wide range of resolutions.}
   {We develop the "anchor-and-refine" stellar feedback implementation. We "anchor" the host cell at each star particle's neighborhood as a specific reservoir for receiving thermal feedback when the cooling radius is resolved, "protected" from being pushed away. If cooling radius can be resolved, the total energy from supernova is injected entirely as thermal energy to the host cell. If not, we refine the host cell and the terminal momentum is injected into the surrounding gas cells.}
   {Idealized tests of our model agrees well with high-resolution numerical works. Our model remains stable across mass resolutions ($10^1$ to $10^5 \; \rm M_{\odot}$). The model maintains consistent superbubble properties (thermal energy, physical radius and radial momentum), and the hot gas interior is preserved without artificial cooling. In the isolated Milky Way-mass galaxy run, the "anchor-and-refine" scheme provides more fine X-ray emitting structures and reveals a sharply defined X-ray bright central mid-plane. The star formation history under the "anchor-and-refine" scheme is smoother than the pure hybrid feedback (the averaged burstiness of star formation, $\delta = {\rm SFR - \langle SFR \rangle}$, are 0.23 and 0.69 $\rm M_\odot / yr$ respectively). Moreover, hot gas dominates mass loading in outflows, with a mass-loading factor plateauing at around 0.2 for heights z > 1 kpc. Decomposition of outflow phases reveals that hot gas drives a steady, energy-driven wind, while cooler phases decline with height due to radiative cooling and gravitational fallback. This highlights the critical role of thermal feedback in galactic winds.}
   {The "anchor-and-refine" scheme successfully maintains consistent superbubble properties, and the hot gas interior is self-consistently preserved, ensuring accurate energy partitioning between thermal and kinetic components. It has significant advantage in capturing hot component of fine structures and regulating star formation and feedback activities for galactic simulations. The robustness of the model is crucial for applying the model to galaxy-scale and zoom-in cosmological simulations where resolution is often limited.}


   \maketitle
\section{Introduction}\label{sec:intro}

Stellar feedback plays a critical role in the evolution of galaxies. The absence of stellar feedback would result in excessive cooling of the interstellar gas, which would then be converted too efficiently into stars resulting  in stellar masses much larger than the observed values \citep[the so-called overcooling problem; for recent reviews see, e.g.,][]{2017ARA&A..55...59N, 2020NatRP...2...42V}. Stellar feedback, mainly from massive stars, injects energy and momentum into the surrounding gas and regulates the phases of the interstellar medium (ISM). Among possible stellar feedback channels, supernovae and radiation are the most widely studied \citep[e.g.,][]{2011MNRAS.417..950H, 2013MNRAS.428..129S, 2013ApJ...770...25A, 2014MNRAS.439.2990S, 2016MNRAS.462.4532G, 2019MNRAS.490.1961E, 2021MNRAS.500.1833S, 2022MNRAS.510.5592A, 2023MNRAS.523.6336B, 2025arXiv251002432Z}.
Stellar feedback can transport metal-rich gas to the circumgalactic medium (CGM) or even the intergalactic medium (IGM) through supernova driven galactic winds, and is therefore an important part of the baryon cycle in galactic ecosystems \citep[e.g.,][]{2014Natur.509..177V, 2014MNRAS.444.1518V, 2017ARAA..55..389T, 2018MNRAS.475..676S, 2019ComAC...6....2N, 2025OJAp....8E.153K, 2025MNRAS.544..513M, 2025ApJ...991..170Z}.

The most direct product of stellar feedback is the emergence of a multiphase ISM structure. The hot phase of ISM mainly corresponds to superbubbles and galactic winds. \citet{1977ApJ...218..377W} construct an analytical model for individual self-similar interstellar bubbles, powered by continuous injection of energy from stellar winds. This model is extended by \citet{1988ApJ...324..776M}, who show that successive supernovae can also produce superbubbles and present a very comprehensive scenario of the dynamical structure and evolution of superbubbles. As a complementary approach to analytical models, numerical simulations have also been carried out on isolated superbubbles \citep[e.g.,][]{2014ApJ...794L..21K, 2015ApJ...802...99K, 2017ApJ...834...25K, 2017MNRAS.465.1720Y, 2018MNRAS.481.3325F, 2019MNRAS.490.1961E}. In general, after accounting for radiative losses, superbubbles retain less energy and momentum than predicted by adiabatic theories.
\citet{2017MNRAS.465.1720Y} perform three-dimensional simulations of isolated superbubbles and find that most of the input energy is lost by radiative cooling, and the superbubbles retain less than $10 \%$ of the input energy. Experiments with multiple correlated SNe show momentum per event remains comparable to single-SN cases, highlighting implications for turbulence driving and star formation regulation in ISM models, while underscoring the need for resolved simulations to avoid overcooling artifacts in galactic feedback prescriptions. \citet{2017ApJ...834...25K} perform three-dimensional hydrodynamic simulations of superbubbles driven by multiple supernovae (with time interval between SN events ranges from 0.01 to 0.1 Myr) in a two-phase, cloudy interstellar medium and conclude that the radial momentum injection per SN ranges from $0.7 - 3 \times 10^{5} \, {\rm M_\odot \, km/s}$.
\citet{2019MNRAS.490.1961E} employ high-resolution, one-dimensional hydrodynamic simulations of superbubbles, with a particular focus on the interplay between thermal conduction and radiative cooling at the interface of the hot bubble interior and the cooled shell. By incorporating an effective diffusivity to approximate turbulent mixing (which is not properly modeled in 1D simulations), the authors demonstrate that the cooling efficiency is primarily governed by the ambient interstellar medium density and the efficacy of non-linear mixing, leading to a revised analytic model that updates the \citet{1977ApJ...218..377W} solution by accounting for radiative losses. This revised model is a crucial benchmark to compare with. for our new treatment of stellar feedback proposed in this work.

Successive and concentrated superbubbles can break out of the galactic disk and drive galactic outflows., so the mass-loading of multiphase gas can provide crucial insights into the intrinsic stellar feedback mechanisms operating within galaxies. Recent simulations have demonstrated that the mass-loading factor ($\eta \equiv \dot{M}_{\text{out}}/\text{SFR}$) is not a simple single-valued function but is highly dependent on the local ISM density, the clustering of stellar feedback events, and the precise implementation of subgrid physics \citep[e.g.,][]{2004AA...425..899D, 2013MNRAS.432.1396G, 2014MNRAS.442.3013K, 2017ApJ...841..101L, 2017ApJ...834...25K, 2019MNRAS.489.4233M, 2021MNRAS.500.1833S, 2022MNRAS.510.5592A, 2023MNRAS.523.6336B, 2025arXiv251002432Z}. For instance, \cite{hopkins2014a} and \citet{2021MNRAS.508.2979P} show that efficient cooling in dense gas leads to high mass-loading factors in dwarf galaxies, effectively regulating their star formation, while in massive galaxies, the mass-loading is lower and dominated by hotter gas phases. 

Moreover, studies of multi-phase outflow have highlighted the complexity of galactic winds, suggesting that the mass loading may be distributed differently among the cool, warm, and hot gas phases \citep{Veilleux2005, Tumlinson2011, 2017ApJ...841..101L, 2023MNRAS.523.6336B}. 
Experiments with multiple correlated SNe show momentum per event remains comparable to single-SN cases \citep{2017ApJ...834...25K}. 
The study concludes that for typical galactic disk conditions, the mass-loading factor of hot winds is generally below unity, implying limited hot gas loss relative to star formation, and highlighted the sensitivity of wind loading to local SN clustering and ISM density.
\citet{2018ApJ...853..173K} employ local simulations from the TIGRESS suite to investigate the properties of multiphase galactic outflows -- specifically distinguishing between hot winds and warm fountains -- driven by self-consistent star formation and supernova feedback in a solar neighborhood environment. They analyze horizontally and temporally averaged flows to quantify mass and energy loading factors, revealing that hot gas, formed via SN shocks and exhibiting minimal radiative losses, behaves as a steady, wind with nearly constant mass flux ($\eta \sim$ 0.1) at distances > 1 kpc from the mid-plane. In contrast, the cool medium ($10^4$ K < T < $2 \times 10^4$ K), while possessing higher instantaneous mass loading ($\eta \sim$ 1 near the disk), experiences significant fallback due to insufficient specific energy compared to gravity, characterizing it as a fountain flow rather than a wind. The study emphasizes that accurate wind loading predictions require phase-separated analysis, providing robust constraints for subgrid models in galaxy formation simulations.

Observational constraints on the mass-loading factor have historically been challenging to obtain, but advancements in ultraviolet (UV) and optical absorption-line spectroscopy have enabled direct measurements of gas flows in and around galaxies. Studies of cool gas outflows via tracers like Si II and Mg II have revealed mass-loading factors ranging from $\eta \sim 0.1$ to $10$ or higher, broadly consistent with theoretical expectations \citep[e.g.,][]{Martin1999, Rupke2005, Heckman2015, Vasan2025}.

The treatment of stellar feedback, especially supernova feedback, for large scale simulations has evolved dramatically. In tradition, the effects of feedback usually need some nonphysical implementations, such as delayed cooling \citep[e.g.,][]{2013MNRAS.429.3068T, 2015MNRAS.454...83W}, stochastic heating \citep[e.g.,][]{2015MNRAS.450.1937C}, or introduce two-phase particles to store the thermal energy \citep[e.g.,][]{2006MNRAS.371.1125S, 2014MNRAS.442.3013K}. \citet{2018MNRAS.478..302S} and \citet{2014ApJS..210...14K} give comprehensive comparisons between these implementations.
Recent innovations focus on parsec-scale resolution, revealing feedback-driven mechanisms like cloud disruption, highlighting the need for self-consistent cluster-scale star formation algorithms to capture feedback's role in galactic outflows and CGM enrichment \citep[e.g., \textit{FIRE-2} and \textit{SMUGGLE}, ][]{2018MNRAS.480..800H, 2019MNRAS.489.4233M}. 
The implementation of feedback has been developed towards a more physical way without artifact procedures, such as mechanical and hybrid feedback schemes. In mechanical feedback \citep[e.g.,][]{2012MNRAS.421.3488H, 2018MNRAS.478..302S}, the energy from supernova explosions is coupled to the surrounding gas cells/particles in the form of momentum (or terminal momentum), while a small portion of energy left is injected as thermal energy. One step forward, the hybrid feedback model \citep[e.g.,][]{2018MNRAS.477.1578H} injects thermal energy if cooling radius is resolved and inject terminal momentum if not. \citet{2017ApJ...846..133K} distinguish more detailed situations when the cooling radius is resolved. Comparison between models \citep{2018MNRAS.477.1578H} shows that the hybrid model has great advantage in producing correct momentum evolution at a wide resolution range, while the thermal component is not properly generated in low resolution runs.

We note that both the mechanical and hybrid treatments of feedback poses an issue in Lagrangian fluid structures. Specifically, the momentum from stellar feedback can directly push away the nearby gas, resulting in insufficient gas cells/particles around the stars to resolve the low-density, hot interior. This issue is prominent in galaxy-scale simulations and will physically underestimate the hot phase of gas. When the hot interior of superbubbles is not captured, the resulting simulations lack a persistent, volume-filling hot component. This missing hot phase leads to an under-prediction of the thermal pressure support and an incorrect partitioning of energy between thermal and kinetic forms in the outflow near the galactic disk. Consequently, the mass-loading factors, the spatial distribution of X-ray emitting gas, and the overall thermodynamic state of the CGM might be systematically biased. This inaccuracy directly influences the predicted properties of baryonic cycle: a weakly represented hot phase can overemphasize the role of cooler outflow and fountain-like recycling processes.
To address this shortcoming, in this work, we employed a "anchor-and-refine" method to resolve the hot gas within superbubbles. This new method performs robustly across a wide range of resolutions (from $10^{1}$ to $10^{5} \, \rm{M_\odot}$). The outline of the paper is as follows. 
In Section~\ref{sec:methods}, we describe the numerical implementation of our "anchor-and-refine" model. In Section~\ref{sec:test}, we perform idealized tests of our model in different resolutions, and then apply it to a simulation of an isolated MW-like mass galaxy in Section~\ref{sec:MW}. We discuss in Section~\ref{sec:discussion} and then briefly summarize the work in Section~\ref{sec:summary}.

\section{Numerical implementation}
\label{sec:methods} 

The numerical improvements developed in this work is based on the \textit{SMUGGLE} framework \citep{2019MNRAS.489.4233M},  a comprehensive galaxy formation model with resolved cold/dense interstellar medium, star formation and stellar feedback processes. It has been used in detailed studies of galactic evolution across a wide range of scales, including properties of giant molecular clouds (GMCs) \citep{2020MNRAS.499.5862L, 2025A&A...699A.282N},  formation and evolution young massive clusters \citep[][]{2022MNRAS.514..265L}, black hole fueling \citep[][]{2022MNRAS.517.4752S}, Lyman-$\alpha$ escaping \citep[][]{2022MNRAS.517....1S} and the effect of feedback in dwarf galaxies \citep[][]{2022PhRvL.129s1103B, 2022MNRAS.513.3458B, 2023MNRAS.520..461J}.
In this work, we have systematically updated the implementation of supernova feedback and stellar wind feedback, as described in the subsections below. We note that we also updated the radiative feedback, which is described in Appendix. \ref{sec:test_ion}.

Star formation is treated the same way as in \citet{2019MNRAS.489.4233M}. Star particles are formed in cold and dense ($n_{\rm gas} > 10 \, \rm cm^{-3}$) molecular gas with a rate $\dot M_{\star}$ defined as
\begin{equation}
    \begin{aligned}
       \dot M_{\star} &= \epsilon_{\rm ff} \; \frac{M_{\rm gas}}{t_{\rm ff}},
	\label{eq:sfr}
	\end{aligned}
\end{equation}
where $\epsilon_{\rm ff} = 0.01$ is the local star formation efficiency, $M_{\rm gas}$ is the mass of the gas cell, and $t_{\rm ff} = \sqrt{\frac{3 \pi}{32 G \rho_{\rm gas}}}$ is the free-fall time of the gas cell with density $\rho_{\rm gas}$.

Following \citet{2019MNRAS.489.4233M}, we estimate the number of type II supernova (SNII) $N_{\rm SNII}$ and the ejected mass $M_{\rm SNII}$ during a timestep (from $t$ to $t+\Delta t$) of a star particle (which represents a coeval simple stellar population) by integrating over the Chabrier initial mass function \citep{2001ApJ...554.1274C}. We set the minimum main-sequence mass to 8 $\rm{M_{\odot}}$ and integration upper limit of the IMF to 100 $\rm{M_{\odot}}$. The ejected mass is estimated as $M_{\rm SNIa} = 1.37 \, N_{\rm SNIa} \, M_{\odot}$, in which $N_{\rm SNIa}$ is the number of type Ia supernova (SNIa) in a timestep and is calculated from a power law delay time distribution \citep[DTD, see][for details]{2013MNRAS.436.3031V}.

\begin{figure}
    \centering
        \subfigure{
        \includegraphics[width=0.47\columnwidth]{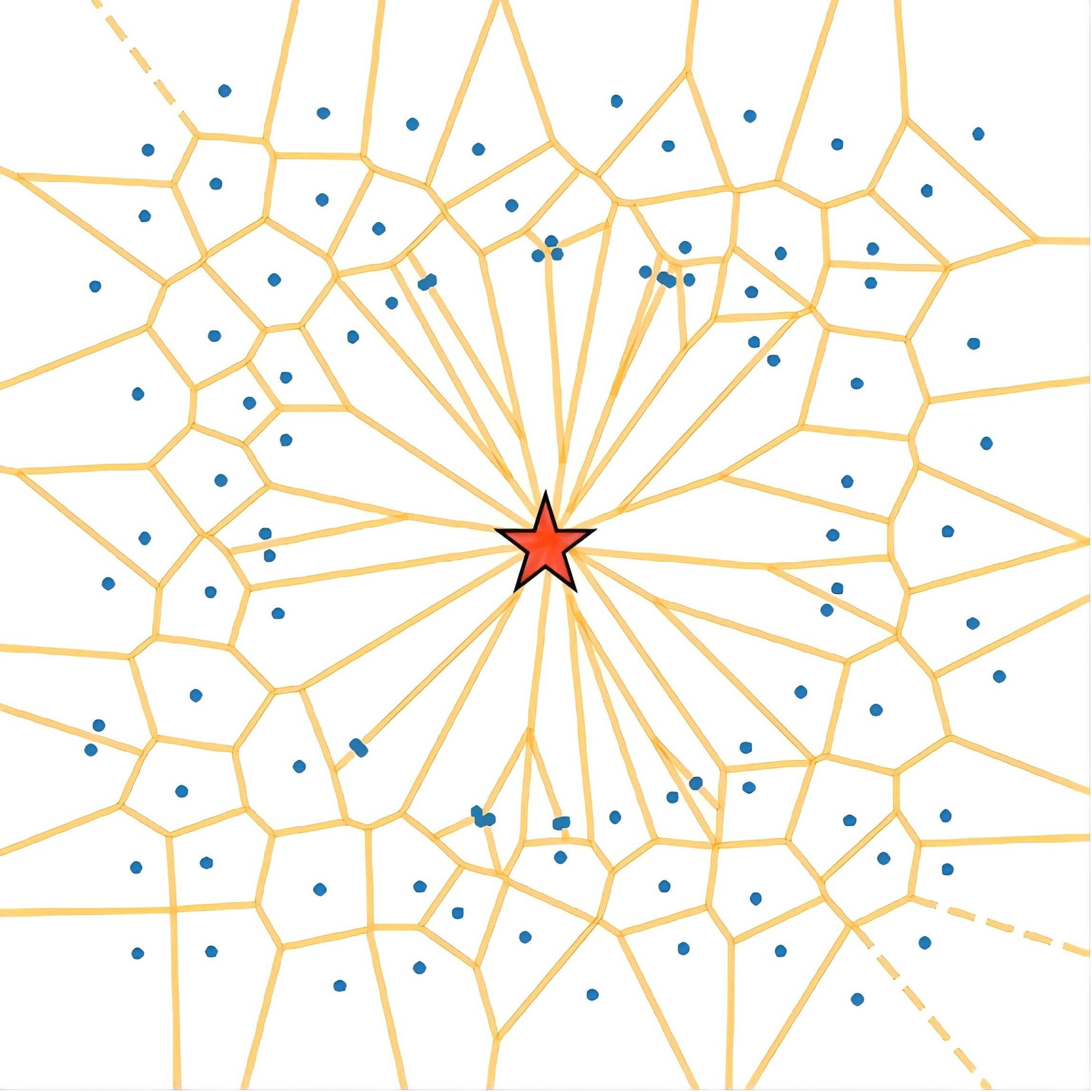}
        }
        \subfigure{
        \includegraphics[width=0.47\columnwidth]{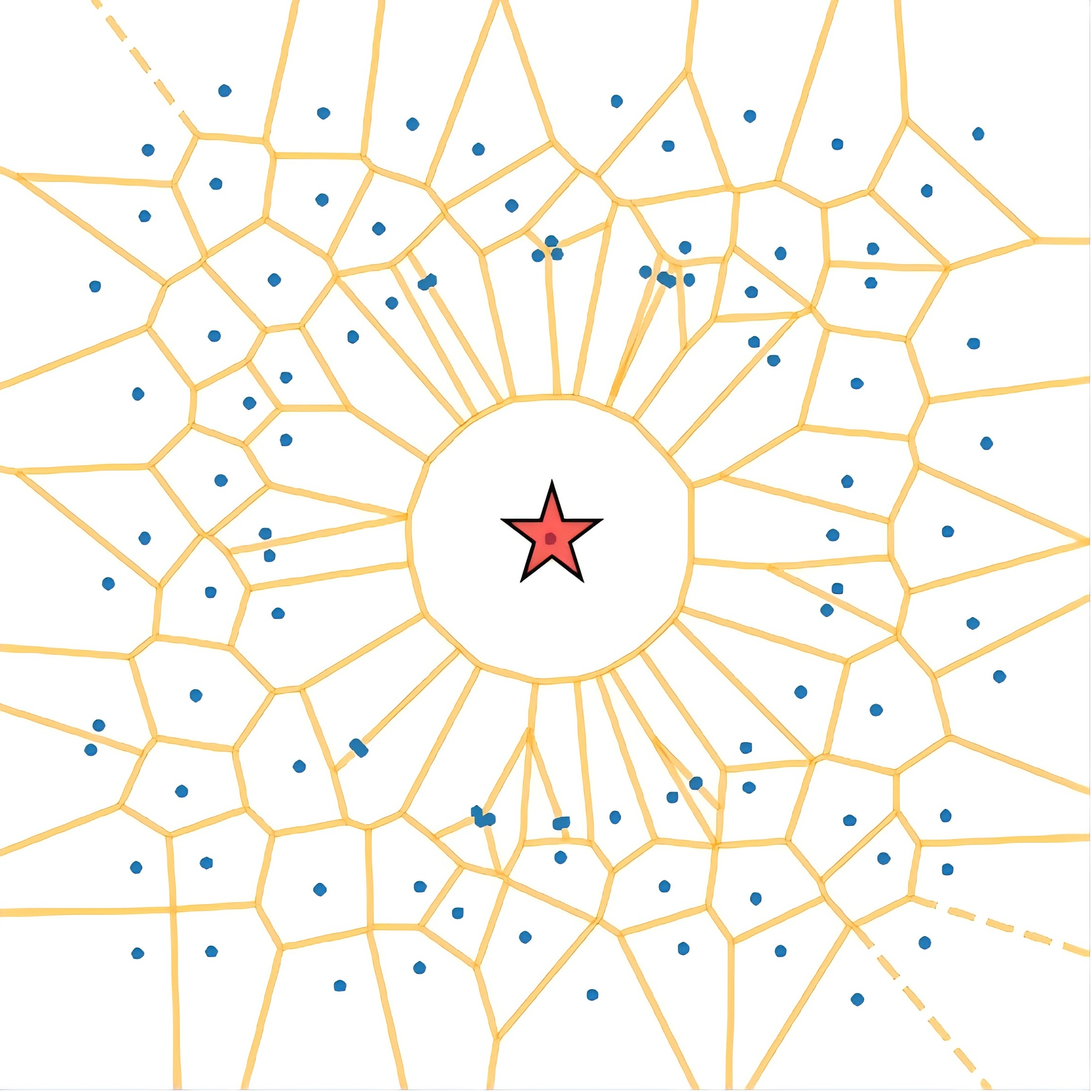}
        }
    \caption{\textit{Left}: An example of a typical mesh structure when supernova feedback is given in the form of momentum. The nearby cells are directly pushed away, resulting in insufficient gas cells around the stars to resolve the low-density, hot interior. \textit{Right}: The sketch of the improved mesh treatment in this work. We "anchor" one (or several) cell(s) at each star particle's neighborhood as a specific reservoir for receiving thermal feedback when the cooling radius is resolved. The cells do not receive any momentum feedback so that they are "protected" from being pushed away.}
    \label{fig:sketch}
\end{figure}

In galaxy-scale simulations usually resolution is not high enough (typically a gas cell/particle has mass $m_{\rm i, gas}> 10^{3} \, \rm{M_\odot}$) to properly resolve the cooling radius (i.e. the shell formation radius), see \citet{2025arXiv251002432Z} for example. In this case, usually the terminal momentum (i.e. the shell formation momentum) is directly injected into the ISM. For the traditional mechanical feedback model \citep[e.g.,][]{2012MNRAS.421.3488H, 2018MNRAS.478..302S}, the energy from supernova explosions is coupled to the surrounding gas cells/particles in the form of terminal momentum, while a small portion of energy left is injected as thermal energy to the surrounding gas cells. One step forward, the hybrid feedback model \citep[e.g.,][]{2018MNRAS.477.1578H, 2017ApJ...846..133K} injects thermal energy if cooling radius is resolved and inject terminal momentum if not.
As mentioned in Sec. \ref{sec:intro}, however, this treatment of feedback poses an issue in (pseudo-)Lagrangian hydrodynamic codes, that the momentum from stellar feedback can directly push away the nearby gas, resulting in an absence of gas cell around the young stars to resolve the low-density, hot gas inside supernova remnants or superbubbles. An example of a typical mesh structure is shown in the left panel of Fig. \ref{fig:sketch}. This issue is prominent in galaxy-scale simulations.

To address this issue, we developed a more effective method to resolve the hot gas within superbubbles at the typical resolution of galaxy-scale simulations. First, we "anchor" the nearest cell in each star particle's neighborhood (i.e. the host cell) as a specific reservoir for receiving thermal feedback when the cooling radius is resolved. It does not mean that the cell is "fixed" in space but, instead, it does not receive any momentum feedback so that it is "protected" from being pushed away. The regularization module in the Arepo code automatically helps to keep the roundness of the cells \citep{2010MNRAS.401..791S}. When a supernova event occurs, a determination is made as to whether the host gas cell of the star particle can resolve the cooling radius. If the cooling radius can be resolved, the total energy $E_{\rm SN} = 10^{51} \rm erg$ released by the supernova is injected entirely as thermal energy to the host gas cell where the star particle resides in; then the hydrodynamics module will naturally capture the evolution of superbubbles described in Appendix \ref{sec:theory} and convert a correct portion of the total energy into kinetic energy via PdV work. If the host gas cell cannot resolve the cooling radius, the host cell is refined into two cells, and then the terminal momentum ($p_{\rm t}$) is injected into the surrounding $N_{\rm ngb} = 64 \pm 2$ gas cells \textit{except for the host cell}. We note that for the simulations analyzed in this work we always take $N_{\rm ngb} = 64 \pm 2$. Specifically, our implementation is described by the following steps:

\textit{1) Estimate primitive coupling radius and find the host cell:} For each star particle, we estimate its primitive coupling radius $h$ through the implicit relation
\begin{equation}
        N_{\rm ngb} 
        = \frac{4 \pi}{3} h^{3} \sum_{i} W(|\boldsymbol{r}_{i} - \boldsymbol{r}_{s}|, h),
	\label{eq:h_coupling}
\end{equation}
where $W$ is the cubic spline SPH kernel \citep{1985A&A...149..135M, 2019MNRAS.489.4233M}, $\boldsymbol{r}_{i}$ and $\boldsymbol{r}_{s}$ are the position vectors of the $i$-th gas neighbor and the star particle respectively. We update $h$ iteratively until the number of neighboring cells enclosed in $h$ is $N_{\rm ngb} = 64 \pm 2$. 
During the iteration, we keep updating the smallest distance $|\boldsymbol{r}_{i} - \boldsymbol{r}_{s}|$ to each star particle and regard the nearest neighboring cell as the host cell.

\textit{2) Mark and anchor the host cell as thermal-energy-receiving cell:} In order to efficiently mark and unmark cells through out the simulation, we adopt a time-based tag. Each cell is assigned a timestamp tag $t_{\rm gas, \, refine}$. Once a star particle identifies its corresponding host cell, it updates the $t_{\rm gas, \, refine}$ value of that cell to the current simulation time $t_{\rm current}$ plus the particle's timestep $\Delta t_{\rm star}$, so that $t_{\rm gas, \, refine} = t_{\rm current} + \Delta t_{\rm star}$. This tag indicates that the cell retains its status as a host cell until the specified time $t_{\rm gas, \, refine}$ is reached. By adopting this approach, we only need to update the tag for newly assigned host cells, without having to explicitly unmark cells that are no longer serving as hosts. Once $t_{\rm gas, \, refine}$ expires, the host-cell status is automatically invalidated. Once a cell is at the status of host cell, it is protected from any momentum feedback injection and is only allowed to receive thermal feedback.

\textit{3) Update coupling radius as the momentum-receiving radius:} After the host cell is protected, the real number of cells enclosed in $h$ which can receive momentum feedback is decreased, which may not be enough to perform momentum feedback anymore. For instance, it is possible that a large fraction of cells in $h$ are being protected in extreme cases like clustered star particles. Thus, this step involves a neighbor-finding procedure similar to the first step, with the key distinction that host cells are no longer considered in the process. Based on the identified momentum-receiving cells, a new coupling radius $h_{p}$ is then updated.

\textit{4) Perform the feedback:} Once a supernova event occurs, the corresponding cooling radius $r_{\rm cooling}$ is estimated as the shell formation radius following \citet{1977ApJ...218..377W}
\begin{equation}
        r_{\rm sf} 
        = 28.4 \; {\rm pc} \; E_{51}^{2/7} n_{0}^{-3/7}, 
	\label{eq:r_sf}
\end{equation}
where $E_{51} = E_{\rm SN} / 10^{51} \rm erg$ and $n_{\rm 0} = n_{\rm amb} / 1 \, \rm cm^{-3}$. $E_{\rm SN}$ is the the injected energy of a supernova and $n_{\rm amb}$ is the ambient density, which is estimated by the density of the host cell. The cooling radius is resolved if $r_{\rm host} < r_{\rm cooling}$, in which $r_{\rm host} = (3 V_{\rm host} / 4 \pi)^{1/3}$ is the effective radius of the host cell and $V_{\rm host}$ is the volume of the host cell.
If $r_{\rm cooling}$ is resolved, we simply inject all of the feedback energy $E_{\rm SN}$ as thermal energy into the host cell. If not, we take two actions. \textit{First}, inject terminal momentum, which follows the scheme described in \citet{2019MNRAS.489.4233M}. We inject total terminal momentum $p_{\rm t}$ into $N_{{\rm ngb}, \, p}$ momentum-receiving neighbor cells found in the coupling radius $h_{p}$:
\begin{equation}
        p_{\rm t} 
        = 4.8 \times 10^{5} \; {\rm M_{\odot} \, km \, s^{-1}} \; 
        E_{51}^{13/14} n_{0}^{-1/7} f(Z),
	\label{eq:p_t}
\end{equation}
where $f(Z) = {\rm min}[(Z / Z_{\odot})^{-0.14}, 2]$ \citep{2019MNRAS.489.4233M}, and the momentum received by each cell is determined by its solid angle weight $p_{i, \rm t} = (\omega_i / \sum_i \, \omega_i) \, p_{\rm t}$, where $\omega_i = \Omega_i / 4 \pi$ and $\Omega_i$ is the solid angle of cell $i$ to the star particle. \textit{Second}, do refinement of the host cell, which split the host cell into two cells. If the host cell remains unable to resolve the cooling radius in subsequent timesteps, this mechanism will keep being triggered, resulting in continuous refinement. However, when the host cell mass is less than $10^{2} \, {\rm M_{\odot}}$ (we choose this value because an injection of a single supernova energy $E_{\rm SN} = 10^{51} \rm erg$ is able to heat a $10^{2} \, {\rm M_{\odot}}$ gas cell onto $\sim 5\times10^{7}$ K), we only inject the terminal momentum without further refining the host cell to prevent computational cost to increase too steeply.

Note that we estimate the stellar wind mass loss of OB and AGB stars $M_{\rm loss}$ from $t$ to $t + \Delta t$ by a function of stellar mass and time following \citet{2018MNRAS.477.1578H}. The total energy of the stellar wind is then $ E_{\rm wind} = L_{\rm wind} \, \Delta t = 10^{12} \, {\rm erg \, g^{-1}} \, M_{\rm loss} \, \Psi$, where $\Psi$ is an empirical function of time \citep{2019MNRAS.489.4233M}.
The feedback algorithm for stellar winds is exactly the same as for supernovae, except that the feedback event occurs every timestep with a different energy $E_{\rm wind}$.

\section{Model tests}
\label{sec:test}

\subsection{SN-driven superbubble}
\label{sec:bubble_SN}

\begin{figure}
    \centering
         \includegraphics[width=\columnwidth]{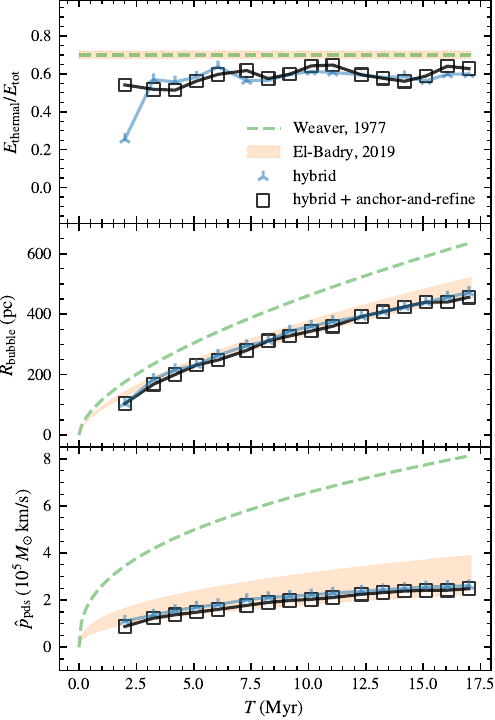}
    \caption{Temporal evolution of a superbubble driven by a \(10^5 \, \rm M_\odot\) star cluster: ratio of thermal to total energy (\textit{top}), physical radius (\textit{middle}), and radial momentum (\textit{bottom}). The blue and black solid lines represent the simulation results of the hybrid feedback and the "anchor-and-refine" scheme (both with mass resolution of 10 $\rm M_\odot$), the green dashed line denotes the classical analytical model \citep{1977ApJ...218..377W}, and the orange shaded area indicates the range of results from the high-resolution 1D simulations in \citet{2019MNRAS.490.1961E}, which depend on assumptions regarding mixing parameters.}
    \label{fig:m1_sum}
\end{figure}

\begin{figure*}
    \centering
         \subfigure{
         \includegraphics[width=0.95\textwidth]{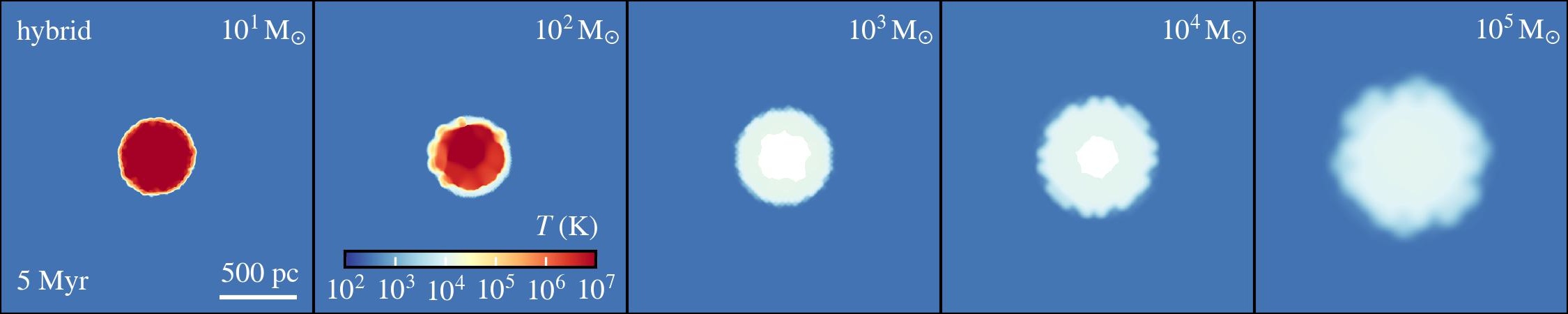}
         }\\
         \subfigure{
         \includegraphics[width=0.95\textwidth]{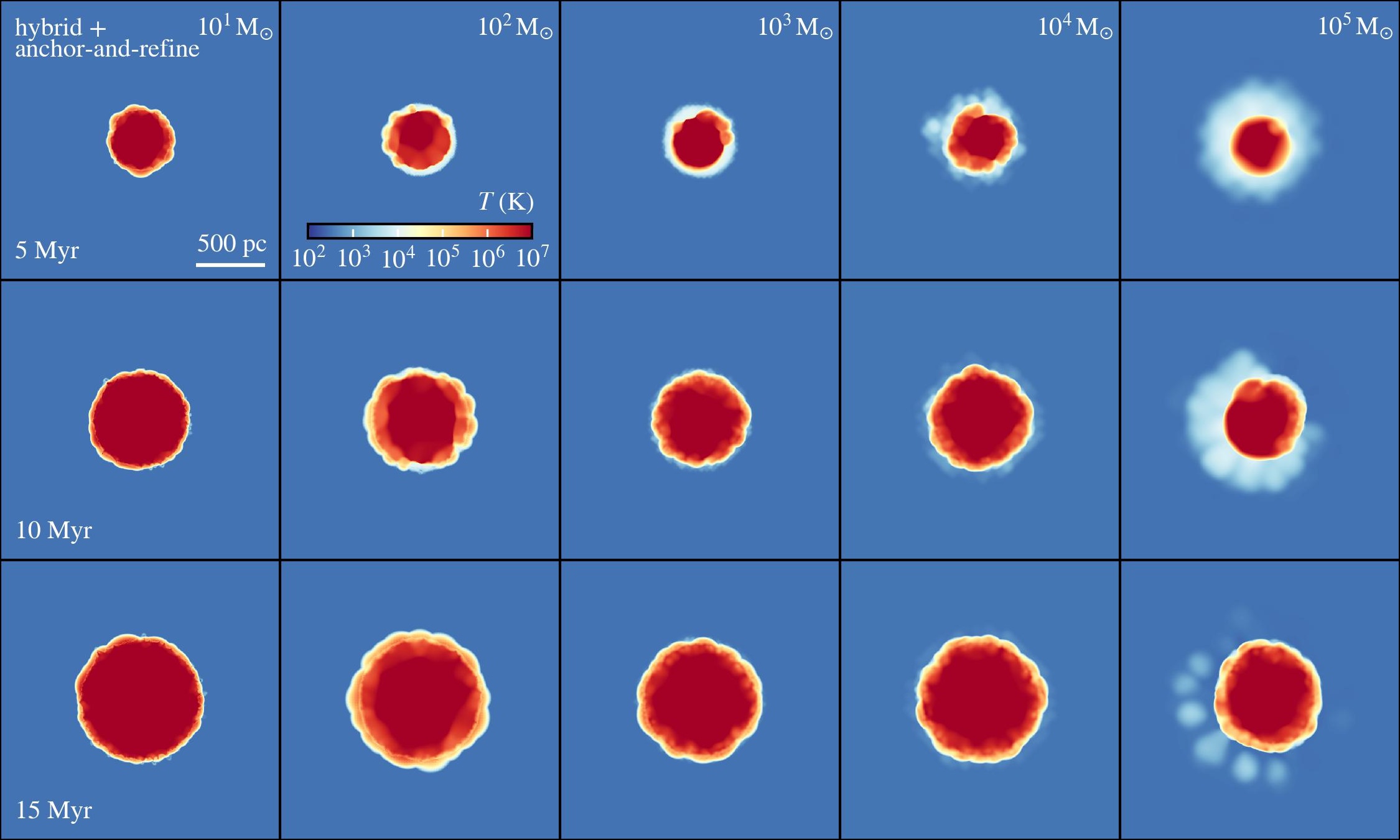}
         }
    \caption{Temperature maps at 5 Myr of a supernova-driven superbubble from a \(10^5 \, \rm M_\odot\) star cluster for the hybrid feedback (\textit{top row}) and the "anchor-and-refine" scheme presented in this work (\textit{bottom rows}, shown at 5, 10, and 15 Myr from top to bottom) for mass resolutions from \(10\) to \(10^5 \, \rm M_\odot\) (left to right). Our model robustly produces a hot bubble interior in a uniform cold environment with density $1 \, \rm{cm^{-3}}$.}
    \label{fig:test_maps}
\end{figure*}

We present a detailed analysis of the evolution of a superbubble undergoing continuous supernova explosions driven by a \(10^5 \, \rm M_\odot\) star cluster in a uniform background with density $n = 1 \, \rm cm^{-3}$ and temperature $T = 100$ K. The calculation of SN events of the star cluster follows the implementation described in Sec.~\ref{sec:methods}. For a star cluster of this mass, the equivalent time interval between successive SN events is roughly $0.02$ Myr, which corresponds to a SN feedback power of  $L_{\rm SN} = 1.54 \times 10^{39} \, \rm{erg/s}$.
Figure \ref{fig:m1_sum} compares the results from two implementations of the \textit{SMUGGLE} model -- both run at a resolution of 10 $\rm M_\odot$ -- namely the "anchor-and-refine" scheme (black sold line) and the hybrid feedback (blue line), against the classical analytical model of \citet[][green dashed line, see also Appendix \ref{sec:theory} for details]{1977ApJ...218..377W}, and the high-resolution 1D numerical results from \citet[][orange band]{2019MNRAS.490.1961E}. At this high resolution, the hybrid feedback yields results consistent with the "anchor-and-refine" model. Overall, both models demonstrate robust performance, with results that agree well with \citet{2019MNRAS.490.1961E}. 

The top panel displays the ratio of thermal energy to total energy, $E_{\mathrm{thermal}}/E_{\mathrm{tot}}$. Although the classical model neglects radiative cooling and thus overestimates the total energy, the partitioning of energy between thermal and kinetic components through PdV work is largely unaffected in principle, leaving the ratio $E_{\mathrm{thermal}}/E_{\mathrm{tot}}$ relatively unchanged. However, our result shows a slightly lower $E_{\mathrm{thermal}}/E_{\mathrm{tot}}$ compared to both \citet{2019MNRAS.490.1961E} and \citet{1977ApJ...218..377W}. This discrepancy may arise from inherent differences between 1D and 3D simulations: the 1D approach requires parameterized mixing assumptions to approximate 3D turbulent dissipation and energy cascades, potentially overestimating thermal energy retention, whereas our 3D simulation naturally captures more realistic, anisotropic energy losses and turbulent mixing processes, leading to more sophisticated radiative cooling processes of the mixing layer. However, 1D simulations are much better resolved and it might also be another reason that thermal energy displays a better retention.

The middle panel, showing the bubble radius $R_{\mathrm{bubble}}$ evolution, demonstrates excellent agreement between our results and the \citet{2019MNRAS.490.1961E} ones, both of which fall below the classical \citet{1977ApJ...218..377W} model (i.e. Eq. \ref{eq:R_pds}). The classical model's neglect of cooling results in excessive pressure and over-expansion. The close match with \citet{2019MNRAS.490.1961E} validates the accuracy of our feedback model in capturing the bubble's hydrodynamic expansion. Similarly, the bottom panel illustrates the radial momentum evolution, where our simulation results align well with \citet{2019MNRAS.490.1961E}, though minor differences may be attributed to the more complex and physically comprehensive treatment of turbulence, mixing, and cooling in our 3D framework. Overall, these results robustly verify the reliability of our stellar feedback model, demonstrating its capability to reproduce key structural and dynamical features of superbubbles while also highlighting the importance of three-dimensional physics in modeling energy conversion and loss processes in supernova-driven flows.

\subsection{Resolution test}
\label{sec:resolution}

\begin{figure}
    \centering
         \includegraphics[width=\columnwidth]{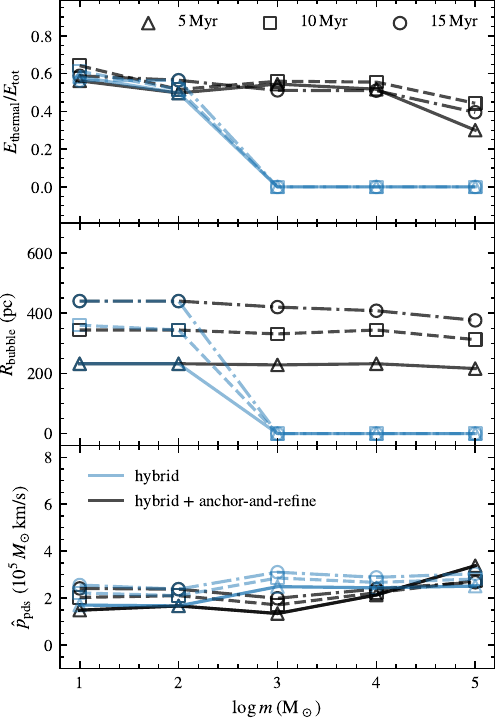}
    \caption{Resolution dependence of superbubble properties driven by a \(10^5 \, \rm M_\odot\) star cluster for the hybrid feedback (\textit{blue}) and the "anchor-and-refine" scheme (\textit{black}). From top to bottom:  ratio of thermal to total energy,  physical radius, and radial momentum plotted against mass resolution. Symbols represent different evolutionary times: 5 Myr (triangles), 10 Myr (squares), and 15 Myr (circles).}
    \label{fig:resolution_test}
\end{figure}

Figure \ref{fig:test_maps} presents numerical tests depicting the temperature distribution of a superbubble generated by a star cluster of $10^5 \, \rm{M_\odot}$ at the center, undergoing successive supernova explosions for the hybrid feedback and the "anchor-and-refine" scheme in this work. We systematically show the temporal evolution  across a range of mass resolutions from $10$ to $10^5$ $\rm M_\odot$. The simulated superbubble exhibits a characteristic structure, with a hot, low-density interior surrounded by a dense, cool shell, which is a hallmark of efficiently expanding supernova-driven bubbles. It is clear to see that the hot component is not present anymore when resolution is worse than $10^3$ $\rm M_\odot$ for the hybrid feedback model.
A key result is the robust performance of our supernova feedback model in terms of structure and morphology of the expanding superbubble across an exceptionally wide range of mass resolutions. Even at the lowest resolution of $10^5 \, \rm M_{\odot}$, the fundamental characteristic of a supernova-driven bubble -- a hot, over-pressurized interior -- is successfully captured. This hot, low-density gas is a direct result of the thermal energy injected by sequential supernovae into the central cell, which naturally prevents rapid radiative cooling and allows for sustained expansion. The superbubble expands symmetrically over time, and the volume filled with hot gas is well-resolved, indicating that the energy injection and its coupling to the interstellar medium (ISM) are effectively modeled. This consistency validates the model's ability to reproduce the fundamental thermodynamics of superbubbles across a significant range in resolution.
It demonstrates that the core physics of energy injection and expansion are correctly captured, ensuring that the model reliably reproduces this dominant feedback mechanism across the resolutions typically employed in galaxy-scale simulations.

However, a deviation is observed at the coarsest resolution of \(10^5 \, \rm M_\odot\). At this scale, the volume of hot gas is reduced, and the radius of the superbubble is slightly smaller compared to the higher-resolution runs, especially at late times (10 and 15 Myr). Although following quantitative analysis of Figure. \ref{fig:resolution_test} shows that the deviation at this resolution from the highest resolution is still acceptable (about 9\% and 15\% at 10 and 15 Myr), we note that the model almost approaches its breaking point. At this limit, the simulation struggles to resolve the internal structure of the superbubble. During the expansion of hot gas, the interaction between the high-res central cells and environmental low-res cells leads to excessive numerical mixing and rapid cooling that artificially suppresses the expansion. This underscores a crucial resolution threshold for our feedback model to reliably capture the thermodynamics and dynamics of superbubbles, whereas coarser resolutions risk underestimation of feedback effects, especially the thermal component.

Figure \ref{fig:resolution_test} presents a comprehensive resolution test of the hybrid feedback (blue lines and symbols) and the "anchor-and-refine" scheme (black lines and symbols), illustrating the evolution of a superbubble driven by multiple supernovae across $10^1$ to $10^5 M_\odot$ resolution range. The top, middle, and bottom panels depict the thermal energy fraction ($E_{\mathrm{thermal}}/E_{\mathrm{total}}$), radius ($R_{\mathrm{bubble}}$), and radial momentum ($\hat{p}$), respectively, with data points corresponding to different evolutionary stages: 5 Myr, 10 Myr, and 15 Myr. A primary result is the stability of the thermal energy fraction (with deviations less than 15\% to the result from the highest resolution) across resolutions higher than $10^5 M_\odot$. For the hybrid feedback, at low resolutions, there is simply no hot gas. However, for the "anchor-and-refine" scheme, the value of $E_{\mathrm{thermal}}/E_{\mathrm{total}}$ remains nearly constant with resolution higher than $10^5 \, \rm M_\odot$, demonstrating that the fundamental physics of energy partitioning -- primarily governed by $\rm PdV$ work during the adiabatic expansion phase -- is correctly captured by our subgrid model. This consistency indicates that the model effectively traps thermal energy within the bubble interior, preventing catastrophic radiative cooling at the coarsest resolutions and validates our feedback implementation as robust and suitable for application in galaxy-scale simulations. 

Similarly, the bubble radius (middle panel) shows minimal variation with resolution. The expansion history is properly reproduced at all resolutions, confirming that the model correctly captures the energy deposition and its coupling to the surrounding medium, leading to a consistent sweep-up of mass and a stable bubble size. At high resolutions ($10^2 \, \rm M_\odot$ or better), the hybrid feedback gives the same results as the "anchor-and-refine" scheme, but there is simply no hot gas at low resolutions.
The radial momentum (bottom panel) also exhibits a high degree of stability, varying by less than a factor of two over the vast resolution range. We can see that even though the hybrid feedback cannot resolve hot component and properly model the expansion of the bubble radius at low resolutions, it can still produce the correct amount of momentum. The slight elevation of radial momentum observed at lower resolutions in our model can be attributed to the longer duration before resolving the cooling radius. During this phase, terminal momentum is injected continuously, leading to a slight cumulative excess compared to higher-resolution runs where thermal injection begins earlier. Overall, the consistency of these key diagnostics demonstrates that our feedback implementation is not overly sensitive to numerical resolution, a critical property for its application in galaxy-scale simulations or even some cosmological simulations.

\section{Application to isolated MW-mass galaxies}
\label{sec:MW}

\begin{figure*}
  \sidecaption
    \centering
         \subfigure{
         \includegraphics[width=12cm]{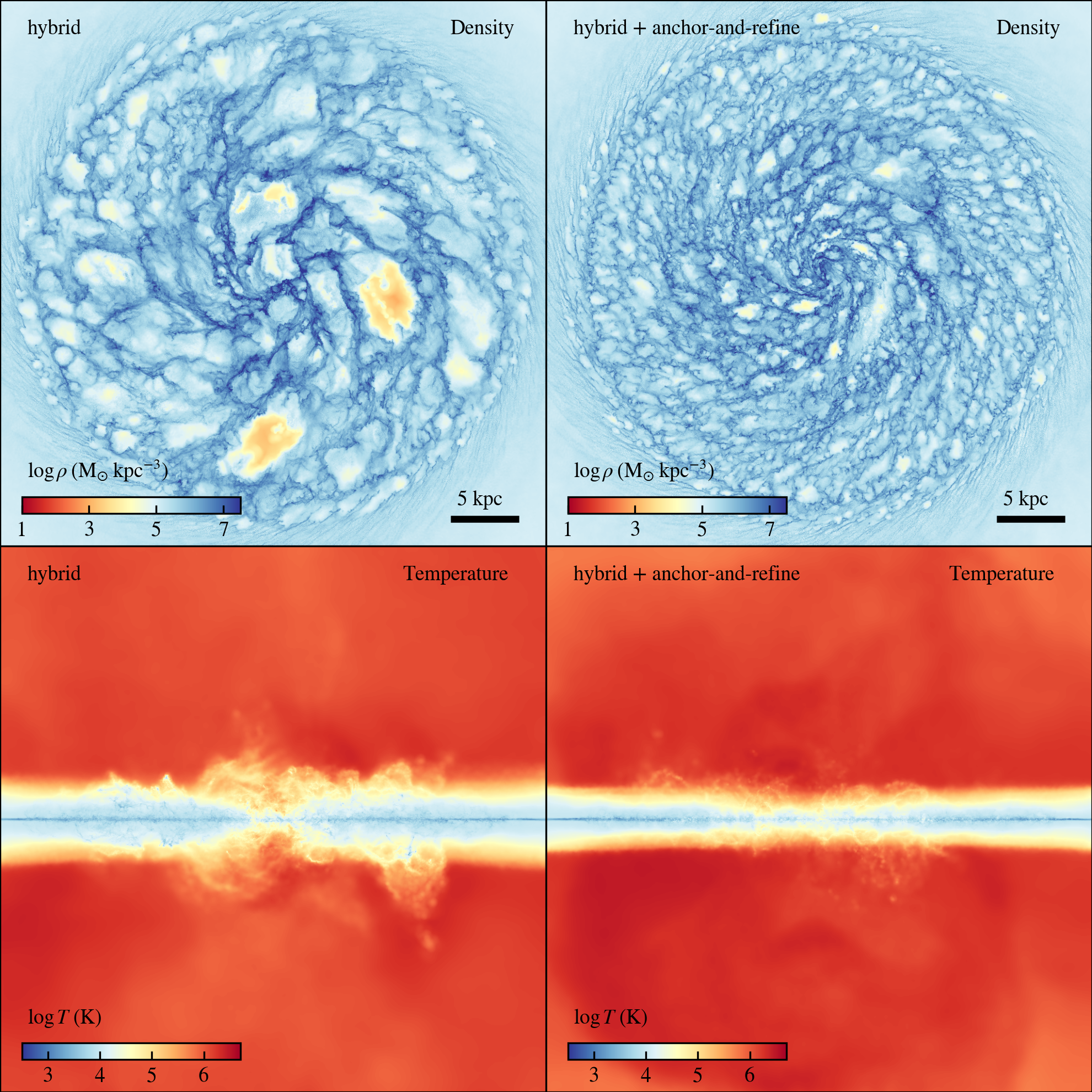}
         }
    \caption{Maps of basic properties of the isolated MW-mass galaxy run for the hybrid feedback (\textit{left}) and the "anchor-and-refine" scheme (\textit{right}). \textit{Top}: face-on density map. \textit{Bottom}: edge-on temperature map. All of the maps are shown at 0.54 Gyr with an extension of $40 \times 40 \, \rm kpc$.}
    \label{fig:maps}
\end{figure*}

\begin{figure*}
  \sidecaption
    \centering
         \subfigure{
         \includegraphics[width=12cm]{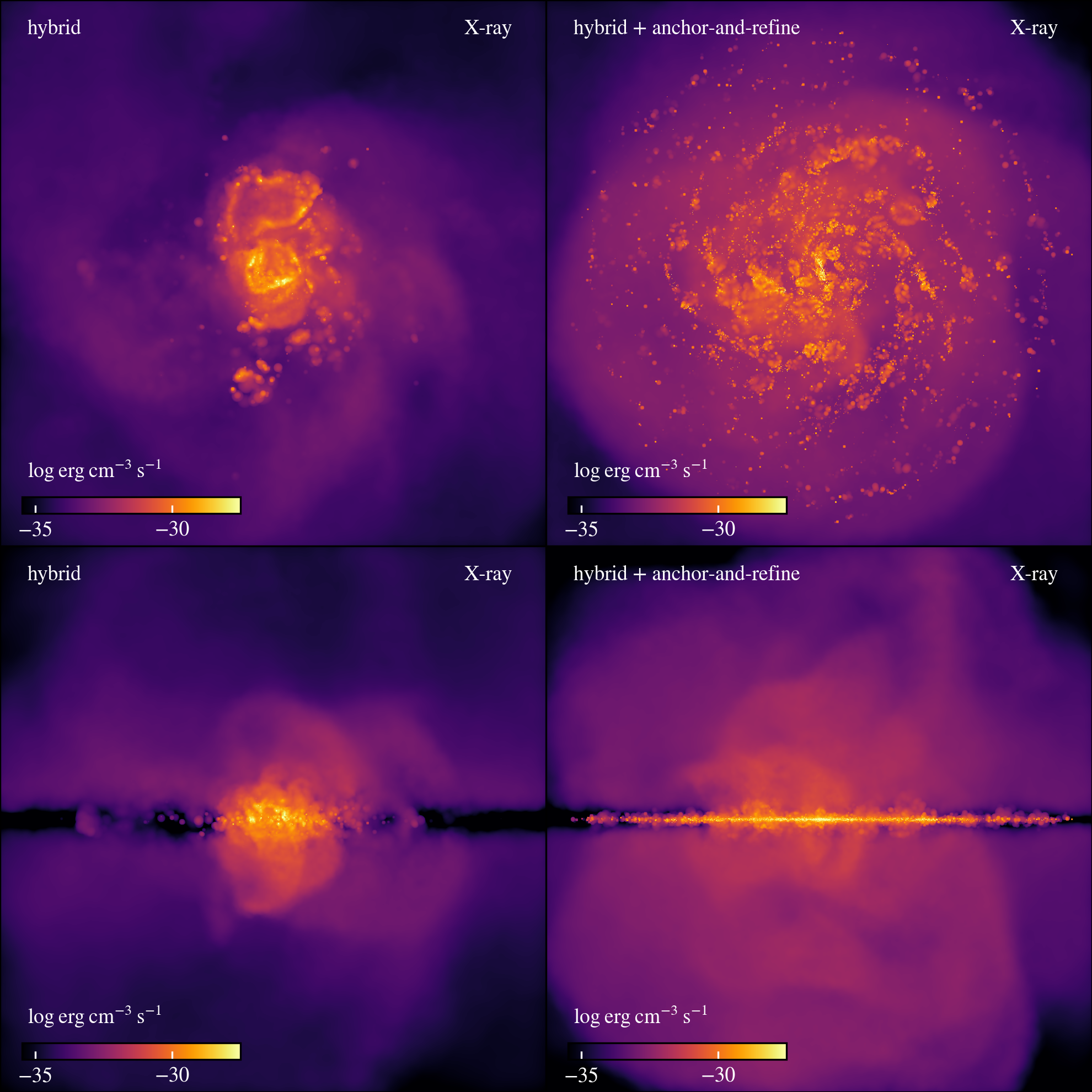}
         }
    \caption{X-ray maps of the isolated MW-mass galaxy run for the hybrid feedback (\textit{left}) and the "anchor-and-refine" scheme (\textit{right}).  The 0.5-2.0 keV soft-band X-ray maps are shown in the face-on (\textit{top}) and edge-on (\textit{bottom}) orientations. The X-ray emission is calculated from the APEC model \citep[][]{2001ApJ...556L..91S}. All of the maps are shown at 0.54 Gyr with an extension of $40 \times 40 \, \rm kpc$.}
    \label{fig:Xmaps}
\end{figure*}

In this section, we conduct an isolated MW-mass galaxy simulation to evaluate the efficacy of our stellar feedback implementation in regulating the interstellar medium and driving outflows.
The initial condition of the galaxy is the same as that in the original \textit{SMUGGLE} work \citep{2019MNRAS.489.4233M}, which has a total mass of $1.6 \times 10^{12} \; \rm M_{\odot}$, with $9 \times 10^{9} \; \rm M_{\odot}$ in stars and $5 \times 10^{10} \; \rm M_{\odot}$ in the gaseous disk. For the gas component, the simulations reach a minimum softening length of $3.6 \, \rm pc$ and a global mass resolution of about $1.4 \times 10^3 \, \rm M_{\sun}$, which are thus able to spatially resolve the structure of superbubbles around stellar associations. In the neighbor regions around star particles, where the "anchor-and-refine" implementation works, the mass resolution can reach values of $\sim 100 \, {\rm M_\odot}$. 

The face-on gas density distribution presented in the top panels of Fig.~\ref{fig:maps} provides some basic review of the structure and dynamics of the interstellar medium regulated by stellar feedback. All the maps are shown at  $t=0.54$ Gyr with an extension of $40 \times 40 \, \rm kpc$. As the hybrid feedback is dominated by momentum in this resolution, the feedback injects mainly kinetic energy, so the voids between filaments are much larger.
The bottom panels of Fig.~\ref{fig:maps} exhibits a characteristic temperature structure in the galactic halo, illuminating the complex interplay between stellar feedback and gas. The central disk ($|z| < \rm{1} \, {kpc}$) maintains cold and cool gas ($T < \rm{10^{4}} \, {K}$, see Sec. \ref{sec:phase} for more details about the multi-phase ISM). Orange filaments transport some cool/warm gas ($T \sim 10^4 - 10^5$ K) out of the disk. A significant difference is shown from comparison between the bottom two panels, that the disk seems to be more disturbed and thicker, and the filaments of cool/warm gas penetrate to larger heights in the pure hybrid feedback run, reflecting its burstier star formation and more kinetic driving of fountain flows.

The soft-band X-ray emission directly quantifies the energy deposition from stellar feedback into the hot phase. The spatial distribution of X-ray emission at 0.5-2.0 keV band in Fig.~\ref{fig:Xmaps} provides an overview of the thermodynamics and kinematics of stellar feedback-regulated gas. 
The face-on view clearly shows complex features where clustered star particles create localized overpressurized regions through feedback. 
We can easily find from the face-on view that the "anchor-and-refine" scheme provides more fine structures. Bright spots exhibit X-ray enhancements of superbubbles or supernova remnants.
We note that the hybrid model can also generate some hot gas in Fig.~\ref{fig:Xmaps}, but is limited in the central region. That is because star formation is highly concentrated in the central region, leading to spatially and temporally correlated supernova events. The collective energy input from many closely packed supernovae can still heat the ambient gas to high temperatures, producing a hot phase in the galactic center. This explains why the hot gas is largely confined to the densest star-forming regions where feedback events overlap; elsewhere in the disk, the thermal component is weak or absent.
The edge-on view reveals a sharply defined central mid-plane with intense X-ray emission for the "anchor-and-refine" scheme. However, the emission from the mid-plane in the hybrid feedback galaxy mainly centered at the inner disk. The bipolar lobes show huge (higher than 10 kpc) outflowing X-ray emitting gas which are reminiscent of the eROSITA Bubbles \citep{2020Natur.588..227P}.

\subsection{Star formation history}
\label{sec:SFH}

\begin{figure}
    \centering
        \subfigure{
        \includegraphics[width=\columnwidth]{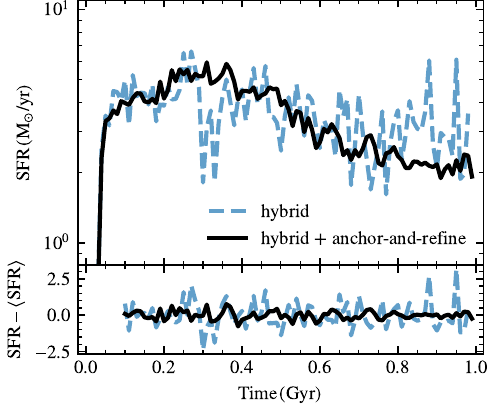}
        }
    \caption{\textit{Top Panel}: Star formation history of the isolated MW-mass galaxy run for the hybrid feedback (\textit{blue}) and the "anchor-and-refine" scheme (\textit{black}). \textit{Bottom Panel}: Burstiness of star formation rate, estimated as $\delta = {\rm SFR - \langle SFR \rangle}$, where $\rm \langle SFR \rangle$ is the star formation history smoothed by a 0.1 Gyr window.}
    \label{fig:SFH}
\end{figure}

Star formation rate and its evolutionary history directly provides information about the possible strength of following stellar feedback. The star formation history directly influences the build-up of stellar mass, the chemical enrichment of the ISM and CGM, and the overall structural evolution of the galaxy – aspects sensitive to how effectively feedback regulates gas cooling, accretion, and the disruption of molecular clouds. The formation history of galaxy offers critical insights into whether the adopted feedback prescriptions realistically capture the complex interplay between stars and their gaseous environment. 

The top panel of Figure \ref{fig:SFH} shows the star formation histories of the isolated MW-mass galaxy under our "anchor-and-refine" implementation and the hybrid feedback. For both models, the galaxy exhibit a rapid rise in SFR during the initial $\sim{0.1}$ Gyr, and declines to about 2 $\rm M_\odot\, yr^{-1}$ after 0.9 Gyr, which is in line with SFR of the MW.
However, the hybrid feedback shows much bursty, where the SFR can vary by nearly an order of magnitude within short intervals. In comparison, the SFR under "anchor-and-refine" scheme are much smoother, which demonstrate that the simulation reaches self-regulation. The bottom panel shows explicitly this kind of burstiness of star formation, which is estimated as $\delta = {\rm SFR - \langle SFR \rangle}$, where $\rm \langle SFR \rangle$ is the star formation history smoothed by a 0.1 Gyr window. The averaged $\delta$ for the "anchor-and-refine" and the hybrid feedback are 0.23 and 0.69 respectively. The reason of the significant burstiness of star formation from the hybrid feedback might be that the pure compression from momentum feedback gives more disturbance to the disk, which results in more giant molecular clouds and more massive star clusters. Also, as we've seen in Fig.~\ref{fig:Xmaps}, the "anchor-and-refine" scheme of feedback resolves much more localized hot bubbles on the disk, which gives a more comprehensive regulation to the star formation activities.

\subsection{Mass loading}
\label{sec:loading}

\begin{figure*}
    \centering
        \subfigure{
        \includegraphics[width=0.48\textwidth]{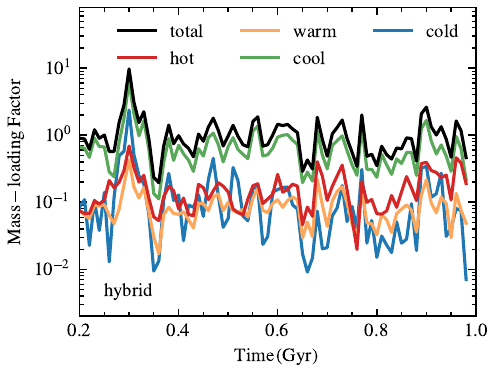}
        }
        \subfigure{
        \includegraphics[width=0.48\textwidth]{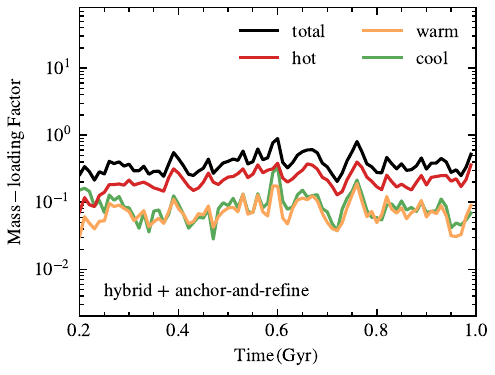}
        }\\
        \subfigure{
        \includegraphics[width=0.48\textwidth]{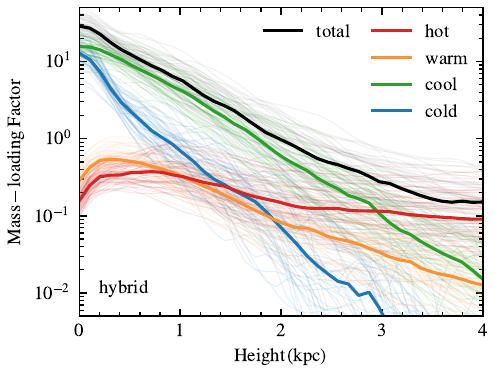}
        }
        \subfigure{
        \includegraphics[width=0.48\textwidth]{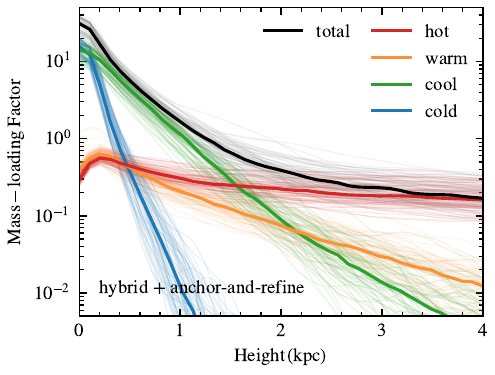}
        }
    \caption{Phases: hot ($T>10^{5.5}$ K, red), warm ($1.7\times10^4 < T < 10^{5.5}$ K, orange), cool ($10^{3.5} < T < 1.7\times10^4$ K, green), cold ($T<10^{3.5}$ K, blue). Total shown in black. \textit{Top}: Time evolution of mass-loading factor ($\eta \equiv \dot{M}_{\text{out}}/\text{SFR}$) in the isolated MW-mass galaxy run for the hybrid feedback (\textit{left}) and the "anchor-and-refine" scheme (\textit{right}), determined by considering the outflowing gas within a galactocentric distance $R \leq$ 40 kpc (projected in the disc plane) contained in two slabs located $\pm$ 2 kpc above and below the mid-plane of the galaxy and 100 pc thick. The total mass-loading is decomposed into different phases (see Sec. \ref{sec:phase} for more details about the multi-phase ISM). Note that the mass loading of cold phase is negligible at this height. \textit{Bottom}: Mass-loading factor versus the distance to the galactic mid-plane for different gas phases. The faint lines represent the scatter from multiple simulation snapshots (from 0.3 to 1.0 Gyr), indicating the variability in the outflow properties. All phases show a decline. Notably, the hot gas $\eta$ plateaus at $0.1$--$1$ for $z > 1$ kpc, indicating a steady-state, energy-driven wind, while the cooler phases continue to decline due to gravitational effects.}
    \label{fig:loading}
\end{figure*}

The top panels of Fig.~\ref{fig:loading} shows the time evolution of mass-loading factor in the isolated MW-mass galaxy run. We consider gas contained in two slabs of thickness 100 pc and located at a height $\pm$ 2 kpc from the mid-plane of the galaxy, within a galactocentric radius $R \leq$ 40 kpc (projected in the disc plane). The mass-loading factor is estimated as $\eta \equiv \dot{M}_{\text{out}}/\text{SFR}$, where $\dot{M}_{\text{out}}$ is the rate of gas mass flowing away from the mid-plane \citep{2025A&A...697A.121B}. The total mass-loading is decomposed into hot ($T > 10^{5.5}$ K, red), warm ($1.7 \times 10^4 \, {\rm K} < T < 10^{5.5} \, {\rm K}$, orange), cool ($10^{3.5} \, {\rm K} < T < 1.7 \times 10^4 \, {\rm K}$, green) and cold ($T < 10^{3.5}$ K, blue) phases (see Appendix \ref{sec:phase} for more details about the multi-phase ISM). Note that the mass loading of cold phase is negligible at this height. The amplitude of mass-loading factor quantifies feedback efficiency in overcoming galactic potential wells, and our model results in an overall mass-loading factor of about $0.1 - 1$ over time, which is in line with \citet{2018ApJ...853..173K}. The oscillations of mass-loading factor indicate the periodicity of star formation and feedback regulation. A detailed decomposition into phases reveals that the hot gas component dominates the total mass-loading factor at this height, underscoring the primary role of thermal injection in generating high-temperature outflows. Comparing the top panels of Fig.~\ref{fig:loading}, we can clearly see that the mass-loading factor of cold and cool gas is significantly higher in the galaxy using the hybrid feedback. The dominant outflowing phase for the "anchor-and-refine" scheme is the hot phase but for the hybrid feedback it is the cool phase. Moreover, the hybrid feedback shows greater oscillations of mass-loading factor, probably due to the more bursty star formation history. 

The bottom panels of Fig.~\ref{fig:loading} presents the mass-loading factor $\eta$, as a function of height above the galactic mid-plane, decomposed by gas phase. The faint curves represent the scatter from multiple simulation snapshots (from 0.3 to 1.0 Gyr), indicating the variability in the outflow properties. A general trend observed across all phases is the decline of $\eta$ with increasing vertical height, $z$. The profile of $\eta_{\mathrm{total}}$ is dominated by the cold and cool gas near the mid-plane but may become increasingly influenced by the persistent hot component at higher altitudes. The steep gradient of the cold and cool phases is a direct consequence of the intense gravitational potential of the galactic plane. They are mainly driven by kinetic energy. When kinetic energy is not strong enough against the gravity, these cold and cool gas finally drop back into the disk as "fountains". It's also consistent with the bottom panels of Fig.~\ref{fig:loading} in which cold and cool gas can reach much higher using the hybrid feedback than using the "anchor-and-refine" method.

The most significant finding is the distinct behavior of the hot gas phase compared to the cooler phases. For $z > 1$ kpc, the mass-loading factor of the hot gas, $\eta_{\mathrm{hot}}$, plateaus at a nearly constant value between approximately $0.1 - 1$ for both models. This asymptotic behavior signifies a transition to a quasi-steady-state outflow for the hot component. The hot gas, primarily heated and accelerated by supernova feedback to temperatures where radiative cooling is inefficient, possesses high thermal pressure and a large scale height. Once it escapes the immediate gravitational confinement of the dense disk ($z > 1$ kpc), the mass-loading factor is roughly constant. It might indicate that the hot gas 
expands almost adiabatically in this regime, and the decrease in density with height is balanced by the increase in the pressure gradient. The plateau value of $\eta_{\mathrm{hot}} \sim 0.1$--$1$ indicates that the feedback processes are generating a sustained, energy-driven wind that carries a mass flux slightly lower than the galaxy's star formation rate into the circumgalactic medium.

The warm phase has a similar mass-loading factor as hot phase but exhibits a continued, steeper decline. It is likely entrained at the interface between the hot wind and the dense ISM, rather than being directly accelerated by thermal pressure. As the outflow rises, this entrained warm gas is prone to being stripped, falling back due to gravity, or mixing and evaporating into the hot phase, leading to a decreasing mass flux with height. The scatter (shown as faint lines) underscores the dynamic and stochastic nature of stellar feedback, where star bursts can cause some variations in the instantaneous outflow properties.

\section{Discussion}
\label{sec:discussion}

The results of Sec. \ref{sec:MW} suggest that reproducing merely the terminal momentum of unresolved supernova remnants may not capture all their galaxy-scale effects; the treatment of the hot superbubble interior also matters. For example, the “anchor-and-refine” scheme reduces the mean burstiness of star formation by a factor of three (see Fig. \ref{fig:SFH}). In the hybrid model, momentum coupled to dense gas produces stronger compression and more spatially correlated star formation, leading to pronounced cycles of gas accumulation and disruption. Retaining the hot phase instead allows energy from clustered supernovae to accumulate and vent through low-density channels, weakening this synchronization. The hybrid model consequently favors intermittent cool and warm fountains, whereas anchor-and-refine produces a more persistent, high-enthalpy wind. The principal difference may therefore lie in the phase structure, specific energy, and recycling time of the outflow rather than its total mass loading. This could affect coronal evolution in Milky-Way-mass halos and baryon and metal retention in dwarfs.

These differences provide useful context for interpreting results from cosmological simulations. In FIRE, burst-driven winds and recurrent recycling—more than half of the ejected gas is subsequently reaccreted—play a central role in the baryon cycle \citep{Muratov2015Gusty,AnglesAlcazar2017Cosmic,Pandya2021Outflows}. The associated “breathing” cycles have been linked to dark-matter cores, stellar migration, size fluctuations, and population gradients in dwarfs \citep{Chan2015FIREHalos,ElBadry2016BreathingFIRE}, while the transition to smoother star formation in Milky-Way progenitors has been associated with inner-CGM virialization and thin-disk formation \citep{Stern2021InnerCGM,Yu2021BurstyThickDisc,Gurvich2023RapidDiscSettling}. Recent studies indicate that the quantitative strength of these effects can depend on feedback coupling: \citet{Hu2019SupernovaWinds} showed that terminal-momentum injection may not fully reproduce the hot wind when the remnant cooling scale is unresolved, while \citet{Zhang2024BurstyDwarfs} found that changing the direction of momentum coupling at fixed total momentum substantially altered dwarf-galaxy burstiness, morphology, and central dark-matter structure. 
Our results therefore complement these earlier findings by suggesting that the amplitude and duty cycle of breathing motions, and their consequences for galaxy structure and baryon cycling, may depend partly on how the collective hot phase is represented. Future CGM observations offer a particularly useful means of distinguishing such implementations because models with similar central galaxies can predict different halo thermodynamics and baryon distributions. Phase-sensitive ion columns and covering fractions from HI through CIV, OVI, and NeVIII, hot-gas tracers such as OVII/OVIII and soft X-ray emission, absorption-line kinematics, and fast-radio-burst dispersion measures should therefore be compared using matched synthetic observations \citep{Mina2021SevenDwarfs,Kelly2022ApostleAuriga}.

Some caveats should be noted. First, our model still cannot fully resolve the hot gas in early stage, especially at low resolution, as the algorithm need a few timesteps to refine the host cell, and in this period we still use terminal momentum as the subgrid model.
Second, it is important to note the limitation of the isolated MW-mass galaxy simulation which is in absence of a cosmological environment. This setup inherently lacks the important process of pristine gas inflow from the cosmic web, which is a key ingredient for sustaining long-term galactic growth in a cosmological context. Our future work will involve zoom-in cosmological simulations to incorporate this essential mechanism and provide a more complete picture of galactic evolution. 
Third, full radiative transfer \citep[like RIGEL, for example][]{2024A&A...691A.231D, 2025A&A...704A.240D} is not implemented in the simulations; the effects of radiative feedback (including photoheating and radiation pressure) are calculated from an effective subgrid model.
Finally, we do not consider AGN feedback \citep[e.g., ][]{2025MNRAS.537..817S} in our simulations.

\section{Summary}
\label{sec:summary}

We introduce a robust, resolution-adaptive, "anchor-and-refine" stellar feedback model, as a part of the \textit{SMUGGLE} framework, that focus on capturing the thermal properties of clustered stellar feedback (supernova and OB stellar winds) in galactic simulations. We also made an update on the photoionization module,in order to have correct ionized regions in extreme situations with highly clustered stellar particles.
We conduct idealized tests and isolated galactic simulations to evaluate the efficacy of this new stellar feedback implementation in regulating the interstellar medium and driving outflows. Below are the key conclusions of this work, structured to highlight the robustness and implications of our model.

\begin{itemize}

\item The "anchor-and-refine" scheme enables us to robustly resolve thermal feedback in superbubbles across an wide range of mass resolutions (from $10$ to $10^5 \, {\rm M_\odot}$). The model maintains consistent superbubble properties, and the hot gas interior is self-consistently preserved, ensuring accurate energy partitioning between thermal and kinetic components. This robustness is crucial for applying the model to galaxy-scale and zoom-in cosmological simulations where resolution is often limited.

\item In idealized tests, our model agrees well on thermal energy, physical radius and radial momentum with the high-resolution 1D numerical results from \citet{2019MNRAS.490.1961E}, validating the accuracy of our feedback model in capturing the bubble's hydrodynamic expansion, with minor differences may be attributed to the more complex and physically comprehensive treatment of turbulence, mixing, and cooling in our 3D framework. 

\item The photoionization feedback module performs robustly across resolutions, accurately reproducing Str\"omgren spheres without numerical artifacts. Tests in dense environments show that the iterative photoionization algorithm resolves ionized regions consistent with theoretical predictions, improving the original \textit{SMUGGLE} implementations that may underestimated ionization fronts in extreme situations with highly clustered stellar particles.

\item In isolated Milky Way-mass galaxy simulation, the star formation history under "anchor-and-refine" scheme are smoother than the pure hybrid feedback. The averaged burstiness of star formation, $\delta = {\rm SFR - \langle SFR \rangle}$, are 0.23 and 0.69 $\rm M_\odot / yr$ respectively.

\item The spatial distribution of X-ray emission shows that the "anchor-and-refine" scheme provides more fine structures. 
The edge-on view reveals a sharply defined central mid-plane with intense X-ray emission. However, the emission from the mid-plane in the hybrid feedback galaxy mainly centered at the inner disk.

\item The star formation history is smoother in the "anchor-and-refine" scheme, with burstiness about one third of that in the hybrid feedback simulation.

\item In the "anchor-and-refine" scheme, hot gas dominates mass loading in outflows, with a mass-loading factor plateauing at around 0.2 for heights z > 1 kpc. Decomposition of outflow phases reveals that hot gas drives a steady, energy-driven wind, while cooler phases decline with height due to radiative cooling and gravitational fallback. Moreover, the hybrid feedback shows greater oscillations of mass-loading factor, which is about three times that in the "anchor-and-refine" scheme, probably due to the more bursty star formation history.

\end{itemize}

The "anchor-and-refine" method we develop in this work is designed to robustly capture the thermal component of feedback, primarily from supernovae and stellar winds, in galaxy-scale simulations. It also has potential for future applications in cosmological zoom-in simulations, where it can ensure consistent treatment of thermal feedback from dwarf to massive galaxies despite varying local resolution. Furthermore, our method provides an improved framework of feedback coupling for Lagrangian codes. The core "anchor-and-refine" logic might also be extended to other feedback channels, such as active galactic nucleus (AGN) feedback, where resolving a stable energy injection cavity is equally crucial. In the future, further development of this work may focus on optimizing the algorithm's computational efficiency for large-scale simulations and integrating it with radiative transfer to create a more comprehensive and physics-rich feedback model for the next generation of galaxy formation simulations.

\begin{acknowledgements}
We thank Volker Springel for allowing us access to the Arepo code. This work was supported in part by the National Natural Science Foundation of China through Grant No. 11821303 \& 12373006, by the Ministry of Science and Technology of China through Grant 2018YFA0404502 and 2023YFB3002502, and by China Manned Space Program with grant NO. CMS-CSST-2025-A10. We also acknowledge the Tsinghua Astrophysics High-Performance Computing (TAHPC) platform for providing computational and data storage resources that have contributed to this work. LVS is grateful for the hospitality at KITP. This research was supported in part by grant NSF PHY-2309135 to the Kavli Institute for Theoretical Physics (KITP). 

\end{acknowledgements}

\bibliographystyle{aa}
\bibliography{ref}

\begin{appendix}

\section{Analytic model}
\label{sec:theory}

The evolution of a superbubble is usually considered as successive supernova explosions. The total number of supernova can be expressed as
\begin{equation}
        N_{\rm SN}
        = L_{\rm SN} \, t \, / \,E_{\rm SN}
        = 3.15 \; L_{38} \; t_{6} \; E_{51}^{-1},
	\label{eq:N_SN}
\end{equation}
where $E_{51} = E_{\rm SN} / 10^{51} \rm erg$, $L_{38} = L_{\rm SN} / 10^{38} \, \rm erg \, s^{-1}$ and $t_{6} = t / 1 \, \rm Myr$. $E_{\rm SN}$ is the the injected energy of a supernova, $n_{\rm amb}$ is the ambient density and $t$ is the evolution time.

In the pressure-driven snowplow (PDS) stage, the expansion of the superbubble with radius $R_{\rm pds}$ is mainly driven by the thermal pressure of the heated interior gas. The evolution of $R_{\rm pds}$ is \citep{1977ApJ...218..377W}
\begin{equation}
        R_{\rm pds} 
        = 67.1 \; {\rm pc} \; L_{38}^{1/5} \; n_{0}^{-1/5} \; t_{6}^{3/5},
	\label{eq:R_pds}
\end{equation}
where $n_{0} = n_{\rm amb} / 1 \, \rm cm^{-3}$, and $n_{\rm amb}$ is the density of the uniform environment.

The expansion velocity of the superbubble is 
\begin{equation}
        v_{\rm pds} = \frac{d R_{\rm pds}}{d t}
        = 39.42 \; {\rm km/s} \; L_{38}^{1/5} \; n_{0}^{-1/5} \; t_{6}^{-2/5}.
	\label{eq:v_pds}
\end{equation}
The total mass of supperbubble and shell is considered as the total swept-up mass
\begin{equation}
        M_{\rm tot}
        = \frac{4 \pi}{3} R_{\rm pds}^{3} \rho_{\rm amb}
        = 3.61 \times 10^{4} \; M_{\odot} \; L_{38}^{3/5} \; n_{0}^{2/5} \; t_{6}^{9/5},
	\label{eq:M_tot}
\end{equation}
in which $\rho_{\rm amb}$ is like $n_{amb}$ but mass density.
 We then define the normalized swept-up mass $\hat{M}_{\rm tot}$ as $M_{\rm tot}$ divided by the total number of supernova events
\begin{equation}
        \hat{M}_{\rm tot}
        = \frac{M_{\rm tot}}{N_{\rm SN}}
        = 1.15 \times 10^{4} \; M_{\odot} \; L_{38}^{-2/5} \; n_{0}^{2/5} \; t_{6}^{4/5} \; E_{51}.
	\label{eq:M_tot_per_SN}
\end{equation}
Similarly, the total radial momentum and it's normalized value can be estimated as
\begin{equation}
        p_{\rm pds}
        = M_{\rm tot} v_{\rm pds}
        = 1.43 \times 10^{6} \; M_{\odot} \; {\rm km/s} \; L_{38}^{4/5} \; n_{0}^{1/5} \; t_{6}^{7/5},
	\label{eq:p_pds}
\end{equation}
\begin{equation}
        \hat{p}_{\rm pds}
        = \frac{p_{\rm pds}}{N_{\rm SN}}
        = 4.52 \times 10^{5} \; M_{\odot} \; {\rm km/s} \; L_{38}^{-1/5} \; n_{0}^{1/5} \; t_{6}^{2/5} \; E_{51}.
	\label{eq:p_pds_per_SN}
\end{equation}

The kinetic energy of the superbubble in the PDS stage can be considered as the kinetic energy of the shell \citep{1977ApJ...218..377W, 2019MNRAS.490.1961E}
\begin{equation}
        E_{\rm kinetic} 
        = \frac{1}{2} M_{\rm shell} v_{\rm pds}^{2}
        \approx \frac{1}{2} M_{\rm tot} v_{\rm pds}^{2}
        = \frac{2 \pi}{3} \rho_{\rm amb} R_{\rm pds}^{3}
        \dot{R}_{\rm pds}^{2}.
        \label{eq:K_pds_bubble}
\end{equation}
The thermal (internal) energy depends on $P_{\rm bubble}$, and $P_{\rm bubble}$ can be expressed with
\begin{equation}
        P_{\rm bubble} 
        = \rho_{\rm amb} \dot{R}_{\rm pds}^{2}
        + \frac{1}{3} \rho_{\rm amb} R_{\rm pds} \ddot{R}_{\rm pds}.
        \label{eq:P_with_R}
\end{equation}
The thermal energy is 
\begin{equation}
        E_{\rm thermal} 
        = \frac{3}{2} P_{\rm bubble} V_{\rm pds}
        = \frac{2 \pi}{3} \rho_{\rm amb} R_{\rm pds}^{3}
        \left(3 \dot{R}_{\rm pds}^{2} + 
        R_{\rm pds} \ddot{R}_{\rm pds}\right).
        \label{eq:U_pds_bubble}
\end{equation}
Therefore, the ratio of kinetic energy to thermal energy is (note that power-law solution, $R_{\rm pds} \propto t^{\eta}$ and $\eta = 3/5$)
\begin{equation}
        \frac{E_{\rm kinetic}}{E_{\rm thermal}} 
        = \frac{\dot{R}_{\rm pds}^{2}}{3 \dot{R}_{\rm pds}^{2} 
        + R_{\rm pds} \ddot{R}_{\rm pds}}
        = \frac{\eta}{4 \eta - 1},
        \label{eq:KU_ratio}
\end{equation}
which is a constant that depends only on $\eta$. For a superbubble in the PDS stage, $E_{\rm kinetic}/E_{\rm thermal} = 3/7$, and therefore $E_{\rm thermal}/E_{\rm total} = 0.7$, where $E_{\text{total}} = E_{\text{thermal}} + E_{\rm kinetic}$.

\section{Update of the new photoionization module}
\label{sec:test_ion}

\begin{figure*}
    \centering
         \includegraphics[width=0.95\textwidth]{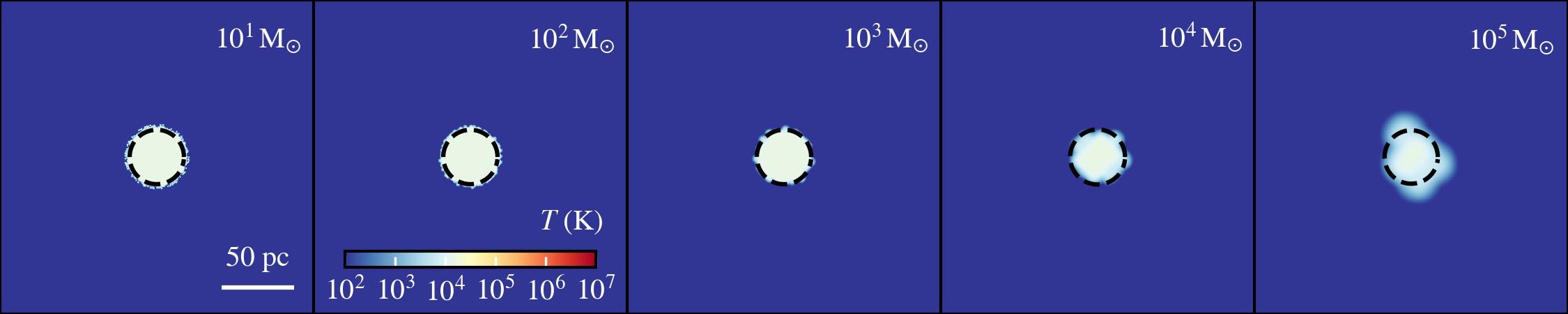}
    \caption{Idealized tests of the new iterative photoionization module. Temperature distribution of gas ionized by a \(10^5 \, \rm M_\odot\) star cluster (taking right after the birth of the star cluster) at different mass resolutions: $10^1 - 10^5  \, \rm M_\odot$ (left to right) in a uniform cold environment with density $100 \, \rm{cm^{-3}}$. The black dashed circle denotes the theoretical  \citep{1939ApJ....89..526S} radius. The photoionization model performs robustly across resolutions, accurately reproducing the ionized region.}
    \label{fig:ion_maps}
\end{figure*}

We consider two kinds of radiative feedback: photoionization and radiation pressure. The photoionization channel is updated from \citet{2019MNRAS.489.4233M} and radiation pressure follows the same implementation as in \citet{2019MNRAS.489.4233M}. After the formation of a star particle, the luminosity of a star particle is estimated with mass-to-light ratio $\gamma_\star$ as
\begin{equation}
        L_\star
        = \gamma_\star \, M_\star,
	\label{eq:L_star}
\end{equation}
where $M_\star$ is the mass of the star particle and we set $\gamma_\star = 10^3 \, {\rm L_\odot / M_\odot}$ by default. For a cell with number density $n_{\rm H}$ and volume $V$, the luminosity need to ionize the cell is
\begin{equation}
        L_{\rm rec}
        = \alpha_{\rm rec} \, n_{\rm H}^2 \, V \, \left < h \nu \right >,
	\label{eq:L_rec}
\end{equation}
where $\left < h \nu \right > = 17 \, \rm eV$ is the average energy of the ionizing photons and $\alpha_{\rm rec} = 2.6 \times 10^{-13} \, \rm cm^3 \, s^{-1}$ is the hydrogen recombination rate \citep{2019MNRAS.489.4233M}. In eq.~\ref{eq:L_rec}, the average hydrogen number density $n_{\rm H}$ is calculated with $n_{\rm H} = X_{\rm p} / m_{\rm p}$, where $X_{\rm p}$ is the hydrogen mass fraction and $m_{\rm p}$ is the proton mass. 

In the original \textit{SMUGGLE} model, each star particle ionize surrounding gas cells in the coupling radius independently. This will underestimate the size photoionized bubbles (the so-called Strömgren radius) when many star particles are gathered in a small region and their photons come out ionizing the same cells multiple times. We update the algorithm with an iterative photoionization search radius which can correctly reproduce the extent of the Strömgren sphere. In each iteration, we perform four steps, "guess-mark-ionize-check", described as follows:

1) \textit{Search for cells in guessed ionizing radius}: At iteration $i$, in this first step we guess a possible ionizing radius $R_{\rm guess, \, i}$ for the star particle and search for gas cells within $R_{\rm guess, \, i}$ one by one. The initial guess of the ionizing radius of a star particle is its Strömgren radius without considering other nearby star particles \citep{1939ApJ....89..526S}
\begin{equation}
        R_{\rm guess, 0} = R_{\rm s}
        = \left ( \frac{3 \, L_\star}{4\pi \alpha_{\rm rec} \, n_{\rm H}^2 \, \left < h \nu \right >} \right )^{1/3}.
	\label{eq:R_s}
\end{equation}

2) \textit{Mark cells feasible for ionization}: For each gas cell obtained in step 1), if it is already ionized or marked by other star particles, skip and continue to the next cell. If not, compare its $L_{\rm rec}$ with the residual luminosity of the star particle $L_{\star, \rm residual}$, which exclude the luminosity already used in ionizing or marking cells from $L_\star$. The probability of ionizing the cell is $p = L_{\star, \rm residual} / L_{\rm rec}$. If $p < 1$, we generate a random number $p'$ with uniform distribution in range [0,1[. if $p' < p$, the cell is marked for phototionization with the star particle's ID, and we update the luminosity of the star particle to 0 as all of the photons from this star particle are used up. If $p >= 1$, the cell is directly marked by the star particle, and then we update the residual luminosity of the star particle as $L_{\star, \rm residual}' = L_{\star, \rm residual} - L_{\rm rec}$. 

3) \textit{Ionize marked cells}: For a gas cell marked as eligible for photoionization, the star particle will ionize it immediately. The cell is then heated to a temperature $T_{\rm ion} = 1.7 \times 10^4 \, {\rm K}$, which corresponding to the temperature of hydrogen ionized by a photon with energy $\left < h \nu \right > = 17 \, \rm eV$. The cooling of the particle is disabled for the duration of the parent stellar particle timestep.
Note that as our simulation runs on multiple MPI tasks, it is possible that several nearby star particles on other tasks also mark the same cell in this iteration $i$. In this case, the mark on the cell will be overwritten by the ID of later star particles. Therefore, at the end of iteration $i$, there are more than one star particles trying to mark and ionize the cell, but the cell can records only one of the star particle's ID, so the ionizing actions from other star particles are "fake".

4) \textit{Check actual consumption}: As mentioned above, some of the ionization in step 3) might be "fake", so after navigating all of the cells in $R_{\rm guess, \, i}$, we need to check the actual luminosity consumed for each star particle. At the end of iteration $i$, we sum up all of the information on every tasks, and for each star particle, we can find those cells ionized and marked by star particle's ID, and the consumed luminosity is $L_{\star, \rm consumed} = \sum L_{\rm rec}$. The residual luminosity for the next iteration ($i+1$) is then $L_{\star, \rm residual} = L_{\star} - L_{\star, \rm consumed}$.
If the photons are used up, then this star particle is done. If not, i.e., $L_{\star, \rm residual} > 0$, we enlarge our guess of ionizing radius by multiplying a factor:
\begin{equation}
        R_{\rm guess, i+1} 
        = R_{\rm guess, i} \times {\rm min}  \left ( 2, \, \frac{L_{\star, \rm residual}}{L_{\star}} \right )^{1/3}.
	\label{eq:R_s_2}
\end{equation}
Normally we expand the ionizing volume by a factor of 2, but when the residual luminosity is small, it is expanded to a minimum size which can use up all of the residual luminosity. After the guessed ionizing radii are updated for every star particles, we move into the next iteration, and keep iterating until all of the star particles use up their photons. We impose a maximum radius of $20 H_{\rm sml}$, where $H_{\rm sml}$ is the smoothing length of the star particle. However, in practice this maximum radius is never reached in our tests. 

Radiation pressure is treated as a momentum feedback and follows the original \textit{SMUGGLE} implementation. The injected momentum is \citep{2019MNRAS.489.4233M}
\begin{equation}
        \Delta p 
        = \frac{L_\star}{c} \, (1+\tau_{\rm IR}) \, \Delta t,
	\label{eq:rad_pressure}
\end{equation}
where $\Delta t$ is the timestep of the star and c is the speed of light. The optical depth of the gas to infrared radiation $\tau_{\rm IR} = \kappa_{\rm IR} \, \Sigma_{\rm gas}$, where $\kappa_{\rm IR}$ is the opacity in the infrared band and $\Sigma_{\rm gas}$ is the gas column density. The momentum is injected into neighbor cells weighted by solid angle, same as the injection of momentum from supernovae in Sec. \ref{sec:methods}.

Previous implementation of photoionization in \textit{SMUGGLE} may underestimate the size of ionized region, caused by the communication issue between star particles on different MPI tasks. Our updated implementation successfully addresses this issue without significantly increasing the computational cost.
Figure~\ref{fig:ion_maps} presents the temperature distribution resulting from the photoionization feedback of a \(10^5 \, \rm M_\odot\) star cluster at the center, simulated across a wide range of mass resolutions from $10^1$ to $10^5 \, \rm  M_\odot$ (left to right). Photoionization occurs promptly after the formation of a star particle, and since stars typically form in high-density molecular clouds, we evaluate the performance of our photoionization model in a dense environment ($n = 100 \, \rm cm^{-3}$). A theoretical estimate of the ionized region size, the so-called \citep{1939ApJ....89..526S} radius, represented by the black dashed circle, provides a benchmark for evaluating the numerical results. Overall, the photoionization feedback model performs robustly across all resolutions, successfully capturing the formation and extent of the ionized region even at the lowest resolution of $10^5 \, \rm M_\odot$ probed by our test simulations.

At the highest resolution of 10 $\rm M_\odot$ (leftmost panel), the ionized region displays with a sharp and physically consistent boundary between the ionized and neutral phases. The simulated ionized region aligns almost perfectly with the theoretical prediction, indicating that the model accurately treats radiation transport and ionization equilibrium under well-resolved conditions. As the resolution decreases (moving rightward through the panels), the boundary of the ionized region becomes progressively more diffuse due to numerical smoothing; however, the overall size and morphology remain in good agreement with the theoretical expectation except at $10^5 M_\odot$, where the boundary is diffuse and the spherical morphology is degraded due to numerical smoothing. This demonstrates that the model maintains numerical stability and physical fidelity even when individual cell masses are large. Minor deviations in the lowest-resolution run likely arise from increased numerical diffusion. These results validate the photoionization model as an effective and numerically reliable subgrid prescription for this stellar feedback process in galaxy formation simulations.

\section{Multi-phase ISM}
\label{sec:phase}

\begin{figure}
	\includegraphics[width=\columnwidth]{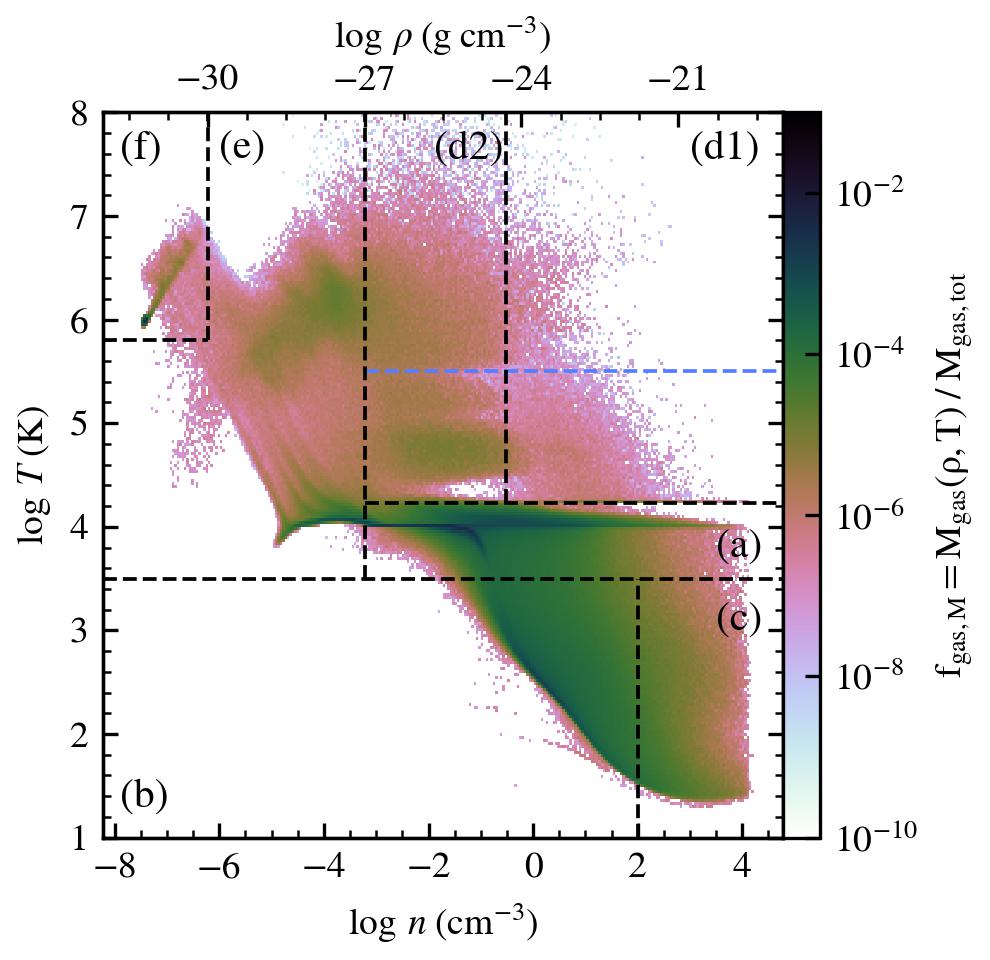}
    \caption{Phase diagram of gas for the isolated MW-mass galaxy run at 0.4 Gyr in the form of a two-dimensional histogram in density and temperature space, with the color of each bin represents the mass fraction of gas at the corresponding density-temperature phase. Different phases are presented as different regions separated by dashed black lines.}
    \label{fig:phase}
\end{figure}

Our model is capable to self-consistently generate a multi-phase structure of the ISM. Fig.~\ref{fig:phase} shows the phase diagram of temperature and density in the isolated galaxy run at 0.4 Gyr.
The diagram is divided into multiple regions, largely following the approach of \citet{2019MNRAS.489.4233M}, with some necessary modifications made according to our study of superbubbles. 
We consider the interstellar medium (ISM) as the gas within the galactic disk, corresponding to phase (a) to (d). The prominent horizontal band of high gas mass fraction retains a warm state as phase (a), with a temperature of about $10^{4} ~\mathrm{K}$. Its high mass fraction indicates efficient radiative cooling has condensed a significant reservoir of diffuse gas. Some of the gas will cool down into cold phase, collapse into dense clumps and form giant molecular clouds (GMCs) in phase (b), with some molecular cores evolve dense enough (i.e., $n > 100 \, \rm{} cm^{-3}$) and become star forming gas in phase (c). Phase (d1) and (d2) both represent the gas emerging from stellar feedback, with (d1) resulting mainly from radiative and stellar wind feedback while (d2) is dominated by supernova feedback. After stars are formed, radiation from stars may ionize the surrounding gas as early feedback, heat the gas to a temperature of about $1.7 \times 10^4 ~\mathrm{K}$ and push the gas away through radiation pressure and stellar winds. This gas primarily occupies phase (d1). The gas in phase (d2) will be even hotter and more diffuse due to supernova events. We note that the warm gas (with temperature lower than $10^{5.5} \; \rm K$, indicated by the dashed blue line) in phase (d1) and (d2) largely corresponds to the observed \ion{H}{II} regions. 

Gas in phase (d2) may eventually flow out of the galaxy and into the circumgalactic medium (CGM), which is largely referred to as phase (e). The outflowing gas will undergo an expansion process, gradually decreasing the temperature and becoming more diffuse. 
Intense stellar feedback in the galactic center can occasionally generate a massive outflow and inject the gas to distances comparable to, or even larger than, the virial radius. Some of the outflow gas may become too distant from the galaxy and cannot fall back, instead it will gradually mix with the gas in phase (f), which represents the background gas environment.
It is important to note that the simulations we employ do not account for the cosmological environment and there is no gas inflow from the IGM, so the CGM consists of only the outflow gas from the ISM and it might be considered as a lower limit to the presence of hot and diffuse gas.

\end{appendix}

\end{document}